# Frustrated Hopping from Orbital Decoration of a Primitive Two-Dimensional Lattice

Aravind Devarakonda,[1,†,‡,*] Christie S. Koay,[2,†] Daniel G. Chica,[2,†] Morgan Thinel,[2] Asish K. Kundu,[3,6] Zhi Lin,[4,§] Alexandru B. Georgescu,[5,¶] Sebastian Rossi,[1] Sae Young Han,[2] Michael E. Ziebel,[2] Madisen A. Holbrook,[1] Anil Rajapitamahuni,[6] Elio Vescovo,[6] K. Watanabe,[7] T. Taniguchi,[8] Milan Delor,[2] Xiaoyang Zhu,[2] Abhay N. Pasupathy,[1,3,*] Raquel Queiroz,[1,*] Cory R. Dean,[1,*] and Xavier Roy[2,*]

[1]Department of Physics, Columbia University, New York, NY, USA

[2]Department of Chemistry, Columbia University, New York, NY, USA

[3]Condensed Matter Physics and Materials Science Division, Brookhaven National Laboratory, Upton, NY, USA

[4]Department of Applied Physics and Applied Mathematics, Columbia University, New York, NY, USA

[5]Department of Materials Science and Engineering, Northwestern University, Evanston, IL, USA

[6]National Synchrotron Light Source II, Brookhaven National Laboratory, Upton, New York, USA

[7]Research Center for Electronic and Optical Materials, National Institute of Materials Science, 1-1 Namiki, Tsukuba, Japan

[8]Research Center for Materials Nanoarchitectonics, National Institute of Materials Science, 1-1 Namiki, Tsukuba, Japan

[†]These authors contributed equally

[‡]Present address: Department of Applied Physics and Applied Mathematics, Columbia University, New York, NY, USA

[§]Present address: Department of Physics, Boston University, Boston, MA, USA

[¶]Present address: Department of Chemistry, Indiana University, Bloomington, IN, USA

[*]Correspondence to: aravind.devarakonda@columbia.edu, apn2108@columbia.edu, rq2179@columbia.edu, cd2478@columbia.edu, xr2114@columbia.edu

**Materials hosting flat electronic bands are a central focus of condensed matter physics as promising venues for novel electronic ground states. Two-dimensional (2D) geometrically frustrated lattices such as the kagome (1), dice (2), and Lieb (3) lattices are attractive targets in this direction, anticipated to realize perfectly flat bands. Synthesizing these special structures, however, poses a formidable challenge, exemplified by the absence of solid-state materials realizing the dice (2) and Lieb (3) lattices. An alternative route leverages atomic orbitals to create the characteristic electron hopping of geometrically frustrated lattices (4–6). This strategy promises to expand the list of candidate materials to simpler structures, but is yet to be demonstrated experimentally. Here, we report the realization of frustrated hopping in the van der Waals (vdW) intermetallic $Pd_5AlI_2$, emerging from orbital decoration of a primitive square lattice. Using angle-resolved photoemission spectroscopy and quantum oscillations measurements, we demonstrate that the band structure of $Pd_5AlI_2$ includes linear Dirac-like bands intersected at their crossing point by a flat band – the essential characteristics of frustrated hopping in the Lieb and dice lattices (7–9). Moreover, $Pd_5AlI_2$ is exceptionally stable, with the unusual bulk band structure and metallicity persisting in ambient conditions down to the monolayer limit. Our ability to realize an electronic structure characteristic of geometrically frustrated lattices establishes orbital decoration of primitive lattices (4–6) as a new approach towards electronic structures that remain elusive under prevailing lattice-centric searches (10,11).**

Electron hopping on a primitive Bravais lattice, for example the two-dimensional (2D) square lattice (Fig. 1a), results in a dispersive band spanning an energy range proportional to $t$, the hopping energy. In the $t = 0$ limit, electrons are localized to atomic sites and the band becomes perfectly flat. Rare earth materials with spatially compact $f$-orbitals approximate this small $t$ limit, and display heavy fermion behavior when their flat band couples with coexisting, dispersive bands (12). Special lattice geometries obtained by adding atomic sites (*i.e.,* a basis) to primitive lattices and connecting them through a specific bond configuration exhibit geometrically frustrated electron hopping that creates flat bands for arbitrary $t$ (13). Prominent examples are the kagome (1), dice (2), and Lieb lattices (3), with the latter constructed by adding atoms at the midpoint of bonds between primitive square lattice sites (Fig. 1b). Although geometrically frustrated lattices promise to expand the realm of bulk flat bands beyond $f$-orbital systems, only kagome and related pyrochlore lattice flat bands have been found in solid-state materials (14–18).

In principle, frustrated electron hopping can be achieved in the absence of frustrated lattice geometry. Particular atomic orbital configurations, for example, introducing two mutually orthogonal $d$-orbitals to the square lattice, which couple only to the corner sites (Fig. 1c), creates frustrated hopping akin to the Lieb lattice. Correspondingly, this configuration we term the decorated checkerboard hosts a similar electronic structure, with coexisting Dirac-like and flat bands (Fig. 1b) in the absence of geometric frustration (Extended Data Fig. 1c). Frustrated hopping arising from the orbital configuration, of which the decorated checkerboard model is an example, has been articulated theoretically (5,6). Identifying a material realization could realize long-sought electronic structures of complex lattice models that have only been accessed in photonic (7–9) and cold atom (19–21) systems, and remain elusive in solid-state materials.

Here, we demonstrate that the van der Waals (vdW) metal $Pd_5AlI_2$ (22) realizes the decorated checkerboard model, with its characteristic Dirac-like bands and flat band positioned close to the Fermi energy. Synthesized from the high-temperature reaction of Pd, Al, and $I_2$ (Methods), single crystals of $Pd_5AlI_2$ grow as thin square plates with metallic luster (Fig. 1d). The crystal consists of two square layers stacked along the $c$-axis (Fig. 1e) to form a tetragonal unit cell (space group $I4/mmm$, Extended Data Table 1). Each layer contains a PdAl checkerboard lattice (Fig. 1f, grey square) sandwiched between $Pd_2$ square nets (Fig. 1f, blue square), which are together capped by iodine atoms. Schematically, this structure can be considered a dimensional reduction of the hypothetical 3D intermetallic $Pd_3Al$ with the auricupride $Cu_3Au$ structure (23). We construct a tight-binding model for the monolayer electronic structure by considering hopping on the PdAl checkerboard between orthogonal Pd [$d_{xz}$, $d_{yz}$] orbitals and Al $p_z$ orbitals (Fig. 1g, see Methods). With electron hopping between $d_{xz}$ and $d_{yz}$ orbitals mediated by a central $p_z$ orbital, this model for monolayer $Pd_5AlI_2$ is equivalent to the 2D decorated checkerboard (Fig. 1b).

We examine the bulk electronic structure of $Pd_5AlI_2$ using angle-resolved photoemission spectroscopy (ARPES). Figure 2a shows an ARPES $E(k)$ intensity map along the $\bar{\Gamma} - \bar{M} - \bar{X} - \bar{\Gamma}$ path of the surface Brillouin zone (BZ) (Fig. 2a, inset), overlaid with the bulk band structure from density functional theory (DFT) calculations (Fig. 2a, dashed and Methods). We find that the electronic structure around the Fermi energy $E_F$ (Fig. 2a, green) is qualitatively consistent with the decorated checkerboard model near half-filling. Unlike the idealized band structure depicted in Fig. 1c, the flat band is dispersive away from the high-symmetry points of the BZ, which we can capture by including finite next-nearest neighbor coupling (NNN) and inequivalent on-site energies for the $d$- and $p$-orbitals (Extended Data Fig. 3b). Moreover, the $E(k)$ intensity maps show little variation with incoming photon energy $hv$ (Extended Data Figs. 4a – 4c), indicating weak out-of-plane momentum $k_z$ dependence, suggestive of a quasi-2D electronic structure.

For small momentum $\kappa = \langle \kappa_x, \kappa_y \rangle$ away from the M and Γ points, however, the decorated checkerboard electronic structure is captured by an effective Hamiltonian

$$\widehat{H} = \hbar v_F \, \boldsymbol{S} \cdot \boldsymbol{\kappa} + \Delta S_z \tag{1}$$

with Δ the quasiparticle mass, and $\mathbf{S} = \langle S_x, S_y \rangle$ the $S = 1$ spin matrices (Methods). For $\Delta = 0$ the electronic structure from Eq. (1) features linearly dispersing bands intersected at $E = 0$ by a flat band, equivalent to the low $E$ limits of the dice and Lieb lattice models (Methods). For $\Delta \neq 0$, due to inequivalent on-site energies and spin-orbit coupling (SOC) for example, a gap is opened at the crossing point. Unique to the decorated checkerboard is the fact that it hosts two independent copies of Eq. (1), one at M and the other at Γ. It is further noteworthy that this Hamiltonian corresponds to 2D pseudospin $S = 1$ Dirac fermions (DFs) (24,25), which are higher spin counterparts of $S = ½$ Dirac fermions in graphene (26).

Closer examination of $E(k)$ near $\overline{\text{M}}$ by ARPES reveals the characteristic flat band between Dirac-like bands (Fig. 2b, left) anticipated for the $S = 1$ DFs, consistent with the DFT calculated $E(k)$ near M (Fig. 2b, right). The $E_F$ intersects the dispersive branch above the flat band, resulting in a small electron pocket $α_M$ with Fermi wavevector $k_{F,α} = 0.06$ Å$^{-1}$ (Fig. 2b, left). We also extract a large Fermi velocity $v_{F,α} = 9.7 \times 10^5$ m/s for these $α_M$ electrons, comparable to $v_F \approx 1.1 \times 10^6$ m/s of massless $S = ½$ DFs in graphene (27,28), suggesting they are highly mobile. The ARPES $E(k)$ near $\overline{\Gamma}$ similarly reveals $E_F$ intersecting a dispersive band (Fig. 2c, left), here yielding a hole-like pocket $β_Γ$ with Fermi wavevector $k_{F,β} = 0.14$ Å$^{-1}$. Comparison to the DFT calculated electronic structure near Γ (Fig. 2c, right) indicates the anticipated flat band lies above $E_F$. Relative to the M point, we extract a smaller $v_{F,β} = 3.1 \times 10^5$ m/s for the $β_Γ$ hole-like carriers. Broadly, these features of the $α_M$ and $β_Γ$ pockets are consistent with the 2D $S = 1$ DFs anticipated at M and Γ, respectively, from the decorated checkerboard model.

An ARPES Fermi surface (FS) map in the $k_x − k_y$ plane (Fig. 2d) reveals three closed pockets: coexisting with the $α_M$ electron and $β_Γ$ hole pockets, we find a large $γ_M$ electron pocket centered at $\overline{\text{M}}$. The apparent cross-sectional areas of the $α_M$ and $β_Γ$ pocket in the FS map are in nominal agreement with those anticipated (Fig. 2d, magenta circles) from their respective Fermi wavevectors, $k_{F,α}$ and $k_{F,β}$ extracted from the ARPES $E(k)$ spectra (Figs. 2b and 2c). We also find good agreement at $E_F$ between the experimental FS map (Fig. 2d) and the DFT calculated Fermi surface projected into the $k_x − k_y$ plane (Fig. 2e). Paralleling the photon energy dependent $E(k)$ maps (Extended Data Figs. 4a – 4c), a photon energy dependent ARPES FS map shows warped cylindrical contours along the $k_z$ direction (Extended Data Figs. 4d and 4e), characteristic of a quasi-2D electronic structure. Indeed, early work found that interlayer resistivity in Pd$_5$AlI$_2$ is five orders of magnitude larger than in-plane resistivity (22), consistent with our ARPES and DFT results. Turning to the $γ_M$ pocket, we link this to the flat band of the ideal decorated checkerboard model, made dispersive in Pd$_5$AlI$_2$ by next nearest-neighbor and higher order electron hopping (Fig. 2a, arrow). Indeed, introducing only NNN hopping to the decorated checkerboard model already produces a $k_x − k_y$ FS map (Extended Data Fig. 3c) similar to Figs. 2d and 2e. More broadly, we find similar agreement between the measured and calculated electronic structure for $E < E_F$ (Extended Data Figs. 4f – 4k), indicating DFT faithfully captures $Pd_5AlI_2$.

As anticipated from these ARPES results, $Pd_5AlI_2$ is a metal, with longitudinal resistivity $\rho_{xx}(T)$ (Fig. 2f) decreasing linearly with decreasing $T$ before saturating at $T \approx 15$ K; the residual resistivity ratio $RRR = \rho_{xx}(300\text{ K}) / \rho_{xx}(2\text{ K}) \approx 14$ is typical of high-quality quasi-2D metals (29,30). Moreover, the magnetoresistance $MR(H) = (\rho_{xx}(H) - \rho_{xx}(0))/\rho_{xx}(0)$ at $T = 3$ K (Fig. 2f, inset) features Shubnikov de-Haas (SdH) quantum oscillations characteristic of high mobility transport that onset at $\mu_0 H^\star \approx 3$ T. From the $H$-dependent damping of the SdH oscillations we extract a quantum mobility $\mu_q \approx 2{,}411$ cm$^2$/Vs, consistent with an independent mobility estimate from the quadratic $MR$ at low $H$ (Extended Data Figs. 5a and 5c). Fast-Fourier transform (FFT) analysis of the SdH oscillations at $T = 3$ K, $\Delta MR(1/H)$, reveals a prominent peak at a frequency $F \approx 130$ T (Fig. 2g). Tracking the SdH oscillation amplitude at fixed $H$ as a function of temperature $I_{SdH}(T)$ (Fig. 2g, inset points) and fitting to the Lifshitz-Kosevich form (Fig. 2g, inset dashed and Methods), we extract a relatively small effective mass $m^\star = 0.16\, m_0$, consistent with the large Fermi velocity of the $\alpha_M$ pocket $S = 1$ DFs. We also examine the variation of $F$ with magnetic field angle $\theta$, and find further evidence for quasi-2D behavior (Extended Data Figs. 5f and 5g) of the $\alpha_M$ pocket.

Examining high-field Hall resistivity $\rho_{yx}(H)$ (Fig. 2h) and $MR(H)$ (Extended Data Fig. 5e) at $T = 0.5$ K up to $\mu_0 H = 36$ T, reveals no SdH oscillations originating from the other pockets, which could be ascribed to their reduced Fermi velocities. However, the non-linear behavior of $\rho_{yx}(H)$ is characteristic of multi-band transport from coexisting FSs. Given the sign of the Hall slope $d\rho_{yx}(H)/dH$ at low $H$ is determined by the highest $\mu$ carriers (Methods), we can link the $\rho_{yx}(H)$ response in this regime (Fig. 2h, blue shaded) to the presence of high $\mu$ electrons from the small $\alpha_M$ pocket. At large $H$, $d\rho_{yx}(H)/dH$ can be used to extract the total carrier density $n_{tot}$, and from a linear fit to $\rho_{yx}(H > 20\text{ T})$ (Fig. 2h, dashed line) we estimate $n_{tot} \approx 1.1 \times 10^{22}$ cm$^{-3}$, which is consistent with the carrier density inferred from the ARPES Fermi surface map and Luttinger's theorem assuming a quasi-2D electronic structure (Methods).

Overall, the transport behavior is typical of a metal. However, the electronic structure is unusual, captured by the decorated checkerboard model hosting 2D $S = 1$ DFs. This raises $Pd_5AlI_2$ as a unique solid-state platform to explore their long-anticipated, novel behavior spanning super-Klein tunneling in transport (32), new topological phases of matter (33,34), and strongly correlated phases (*e.g.,* superconductivity) with enhanced stability (35,36), which continue to attract interest in the yet to be realized dice and Lieb lattices. It is noteworthy that, even in the presence of NNN hopping, the decorated checkerboard model exhibits a weakly dispersing segment along the X – Γ line (Extended Data Fig. 3b, arrow), apparent in DFT calculations (Fig. 2a, arrow), that could still support correlated phenomena; chemical doping into $Pd_5AlI_2$ could be used to access this regime. The observation of quantum oscillations is particularly promising in this respect, given correlated behavior is often fragile against disorder.

Like other vdW materials, few-layer thick flakes of $Pd_5AlI_2$ can be isolated with standard exfoliation techniques, potentially providing access to the true 2D limit of the decorated checkerboard model. Indeed, we can reach the monolayer limit (half unit cell, see Fig. 1d) using conventional methods. Figure 3a shows a representative atomic force microscopy (AFM) image of one such monolayer, as identified by AFM topography and optical contrast (Extended Data Fig. 6), with lateral dimensions exceeding 10 μm. Exfoliated flakes also appear to be air-stable with no evidence of degradation found in AFM topography, optical contrast, or electrical transport after 60 days of exposure to ambient conditions. In parallel, scanning tunneling microscopy (STM) of the $ab$-plane surface of $Pd_5AlI_2$ exposed by cleaving crystals in ambient conditions (Fig. 3b) routinely reveals atomic-resolution images of its square lattice structure. STM images of the same $ab$-plane surface taken after 14 days of exposure to ambient conditions (Fig. 3b, imaged with a different tip) show the same atomic structure without any signs of degradation (*e.g.,* formation of native oxides). Comparable to the chemical stability of graphite (37) and unlike most metals (38,39), the robustness of $Pd_5AlI_2$ is likely linked to the resistance of Pd to oxidation (40).

Figure 3c shows sheet resistance versus temperature $R_\square(T)$ of transport devices with $l = 1, 2, 4,$ and 7 layers of $Pd_5AlI_2$ (Methods). Like bulk $Pd_5AlI_2$, exfoliated flakes show metallic behavior. However, the $RRR$ decreases with decreasing layer number. Whereas $RRR \approx 10$ in bulk and 7-layer samples, the monolayer shows $RRR \approx 2$ (Fig. 3c, inset) with a weak upturn in $R_\square(T)$ for decreasing $T \lesssim 100$ K (Extended Data Fig. 7a), suggesting suppressed $\mu$ towards the monolayer limit. While the $MR(H)$ of the 7-layer device (Fig. 3d) is qualitatively similar to bulk $Pd_5AlI_2$ (Fig. 2f), $MR(H)$ in the monolayer remains positive but becomes cusp-like at low $H$ (Extended Data Fig. 7b and Methods), an indicator of weak-localization by disorder paralleling thin metallic films with SOC (41). Nevertheless, $\mu$ remains sufficiently large down to $l = 2$ to observe SdH quantum oscillations. An FFT analysis of SdH oscillations from devices with $l = 2, 3, 4,$ and 7 (Fig. 3e and Methods) reveals one prominent peak for each device, all of which are in the range $F = 135 - 140$ T, in nominal agreement with bulk $Pd_5AlI_2$ (Fig. 2g and Extended Data Fig. 5d). The temperature and gate dependence of SdH oscillations in exfoliated $Pd_5AlI_2$ further confirms these oscillations originate from the $\alpha_M$ pocket observed in bulk ARPES (Extended Data Fig. 8 and Methods). In parallel, $l$ dependent DFT calculations from monolayer to 4-layers of $Pd_5AlI_2$ show the band structure is qualitatively similar to the bulk (Extended Data Figs. 3f – 3i). Together, the close parallels in transport response, quantum oscillations, and calculated band structures across bulk and exfoliated devices show that $E_F$ is invariant and the $\alpha_M$ pocket $S = 1$ DFs persist with decreasing $l$, and more broadly indicate that the decorated checkerboard remains a faithful description of down to the monolayer limit.

The suppression of RRR in exfoliated $Pd_5AlI_2$ with decreasing $l$, indicative of suppressed charge carrier mobility, suggests that in the monolayer limit charge carriers are increasingly susceptible to scattering by imperfections such as substrate disorder, processing residues, or surface adsorbates (Fig. 4a). It is plausible that for $l > 1$, parallel transport through multiple layers is less susceptible to substrate and surface scattering, resulting in transport behavior resembling bulk $Pd_5AlI_2$. Quantitatively, we can extract an average mobility $\bar{\mu}(l)$ (Fig. 4b, blue) across the $\alpha_M$, $\beta_\Gamma$, and $\gamma_M$ pockets weighted by their carrier densities (Methods), using the relation $R_\Box(l) = 1/(n_{2D}\, e\, \bar{\mu})$, where $n_{2D} = (n_{tot} \times l \times t)$ is the areal carrier density, $t \approx 1$ nm is the thickness of a $Pd_5AlI_2$ layer, and $n_{tot} \approx 1 \times 10^{22}$ cm$^{-3}$ from high-field $\rho_{yx}(H)$ (Fig. 2h) and ARPES measurements (Methods). From this analysis, we find that $\bar{\mu}(l)$ at $T = 1.5$ K decreases from $\bar{\mu} \approx 103$ cm$^2$/Vs at $l = 7$ to $\bar{\mu} \approx 6$ cm$^2$/Vs in monolayer $Pd_5AlI_2$. The evidence for charge carrier localization in the monolayer (Extended Data Fig. 7) suggests $\bar{\mu}$ will become even smaller at $T < 1.5$ K.

Since the large $\gamma_M$ pocket contributes the most charge carriers to transport (Fig. 2d), $\bar{\mu}$ predominantly reflects the mobility of its electrons. The SdH oscillations in exfoliated $Pd_5AlI_2$, on the other hand, provide an independent measure of the quantum mobility for the $\alpha_M$ pocket $S = 1$ DFs. We compute $\mu_q(l)$ for the $\alpha_M$ pocket (Fig. 4b, red) by fitting the $H$ damping of SdH oscillations at $T = 1.5$ K for various $l$ (Extended Data Fig. 8a), and find it reduces monotonically from $\mu_q \approx 2{,}600$ cm$^2$/Vs at 7 layers to $\mu_q \approx 600$ cm$^2$/Vs in the bilayer device. Given that $\mu_q$ is smaller than the usual transport mobility (42), 600 cm$^2$/Vs represents a lower bound for the $\alpha_M$ pocket $S = 1$ DFs in bilayer $Pd_5AlI_2$, approximately 20× larger than $\bar{\mu}$ at the same thickness. The persistence of transport by the $\alpha_M$ pocket $S = 1$ DFs to the monolayer limit is evidenced by the Hall resistance $R_{yx}(H)$ at $T = 0.4$ K, which we scale by $l$ to directly compare devices with different $l$ (Fig. 4c). While the 7-layer device displays non-linear $R_{yx}(H) \times l$ paralleling bulk $Pd_5AlI_2$ (Fig. 2h), in the monolayer $R_{yx}(H) \times l$ is only weakly non-linear up to $\mu_0 H = 31$ T and can be fit by a two-carrier, electron-hole model (Fig. 4c, inset) consistent with the $\alpha_M$ and $\beta_\Gamma$ pockets (Methods); from this fit we estimate $\mu \approx 70$ cm$^2$/Vs for the $\alpha_M$ pocket in the monolayer (Fig. 4b, open circle). Taken together, the thickness dependent transport response of exfoliated $Pd_5AlI_2$ suggests the $\gamma_M$ pocket electrons become localized by disorder on Si/SiO$_2$ at low $T$, leaving behind transport signatures of the $\alpha_M$ and $\beta_\Gamma$ pocket $S = 1$ DFs.

The air-stable metallicity observed here in exfoliated $Pd_5AlI_2$ is a rarity among vdW materials. Figure 4d shows a survey of exfoliated vdW materials sorted by their maximum $\mu$ and total 2D carrier density per layer $|n_{2D}|$ with air-stable (air-sensitive) materials marked by closed (open) symbols (Extended Data Table 2). The vdW metals with large $|n_{2D}|$ tend to be unstable in air when exfoliated and possess relatively small $\mu$ (Fig. 4d, $H$-NbSe$_2$). Although semimetal few-layer graphite (FLG, $n = 4$) can exhibit extremely high $\mu \approx 100{,}000$ cm$^2$/Vs, important for high-quality vdW heterostructures (43), its low $|n_{2D}|$ can be an impediment (44). Sophisticated synthetic techniques can also be used to obtain encapsulated, stable metal films (45) (Fig. 4d, CHet-Ga), however, exfoliated flakes of $Pd_5AlI_2$ represent an easier path that could lead to a new generation of high quality vdW heterostructures, plasmonic devices (46,47), and biosensing applications (48). More broadly, the $Cu_3Au$ structural motif at the core of $Pd_5AlI_2$ is common among intermetallics (49). This suggests that many compounds, overlooked by previous lattice-centric high-throughput searches, could host frustrated hopping and novel electronic structures without frustrated lattice geometries.

## Acknowledgements

We thank J. L. Pack, M. A. Kapfer, E. J. Telford, S. E. Turkel, and J. G. Checkelsky for fruitful discussions. Research on novel hopping models was conducted as part of the Programmable Quantum Materials, an Energy Frontier Research Center, funded by the US Department of Energy (DOE), Office of Science, Basic Energy Sciences, under award DE-SC0019443 (X.R., C.R.D., A.N.P., R.Q., X.-Y.Z., M. D.). The synthesis and structural characterization of the 2D metal was supported by the NSF MRSEC program through the Center for Precision-Assembled Quantum Materials at Columbia University, award number DMR-2011738 (X.R., C.R.D., A.N.P., X.-Y.Z.). The PPMS used to perform vibrating sample magnetometry and electrical transport measurements was purchased with financial support from the National Science Foundation through a supplement to award DMR-1751949 (X.R.). A.D. acknowledges support from the Simons Foundation Society of Fellows (grant No. 855186). High magnetic field measurements were performed at the National High Magnetic Field Laboratory, supported by the National Science Foundation Cooperative Agreement No. DMR-1644779 and the State of Florida. ARPES measurements used resources at the 21-ID (ESM) beamline of the National Synchrotron Light Source II, a U.S. Department of Energy (DOE) Office of Science User Facility operated for the DOE Office of Science by Brookhaven National Laboratory under Contract No. DE-SC0012704.

## Author contributions

C.S.K. and D.G.C. synthesized and characterized bulk crystals with S.Y.H. supporting with EDX and SEM experiments, and M.E.Z supporting with magnetization measurements. A.D. fabricated nanodevices and performed transport characterization. A.D., C.S.K., and D.G.C. conducted high-field measurements. M.T. performed STM characterization. A.K.K. performed ARPES measurements with A.R. and E.V. supporting. Z.L., A.B.G., A.D., and R.Q. performed theoretical calculations. A.N.P., R.Q., C.R.D., and X.R. supervised the project. A.D., C.R.D., and X.R. wrote the manuscript with input from all authors.

## Competing interests

The authors declare no competing interests.

**Data availability**

Data and codes used in this study are available from the authors upon reasonable request.

# Figures

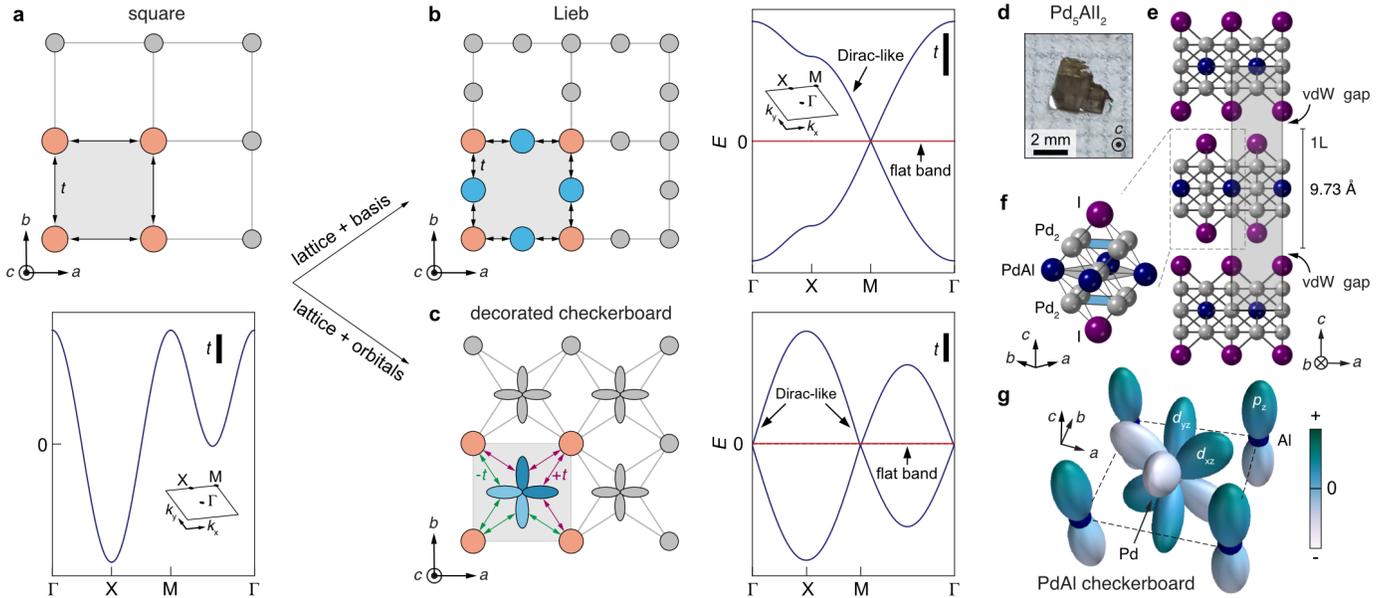

**Figure 1: Frustrated hopping from orbital decoration and the van der Waals metal Pd$_5$AlI$_2$**

**a** (top) Electron hopping on the primitive square lattice, where *t* is the hopping energy, results in (bottom) a single dispersive band in its electronic structure. **b** (top) Introducing a basis to the square lattice by adding sites to the edges of the square unit cell results in the geometrically frustrated Lieb lattice, with (bottom) coexisting Dirac-like and flat bands. **c** (top) Decorating the 2D square lattice with two orthogonal *d*-orbitals (blue) generates frustrated hopping analogous to the Lieb lattice. (bottom) Correspondingly, the resulting electronic is nearly identical with Dirac-like linear band crossings intersected at *E* = 0 by a flat band. Unlike the Lieb, however, direction-dependent sign changes of the hopping energy create two band crossings instead of just one. **d** Optical image of a Pd$_5$AlI$_2$ single crystal. **e** Crystal structure of Pd$_5$AlI$_2$ viewed along the *ab*-plane showing 5-atom thick layers separated by vdW gaps. **f** Monolayer (1L) Pd$_5$AlI$_2$ is composed of a PdAl checkerboard plane (grey square) sandwiched by Pd$_2$ planes (blue square) and capped by I. **g** The PdAl plane with $d_{xz}$ and $d_{yz}$ on the central Pd atom and Al $p_z$ orbitals at the corners maps onto the decorated checkerboard model.

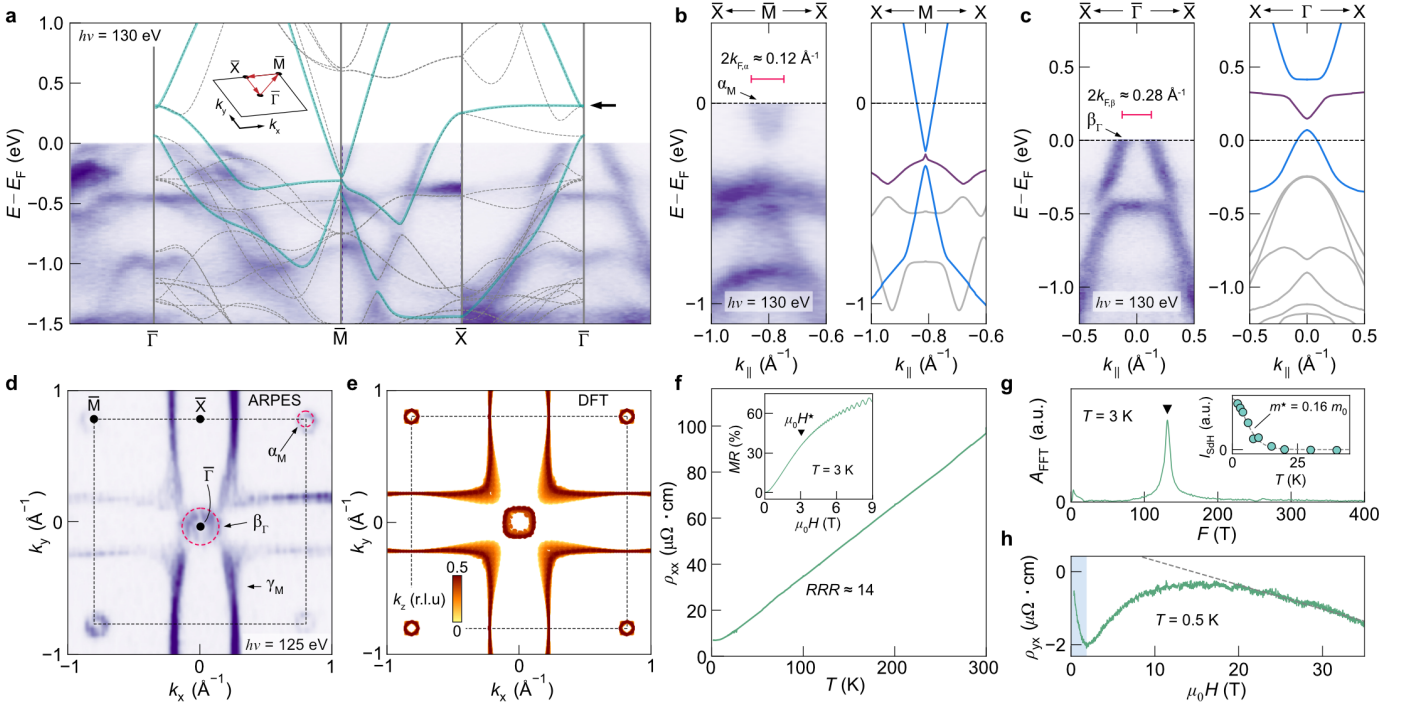

**Figure 2: Bulk electronic structure and transport behavior of Pd$_5$AlI$_2$**

**a** Bulk ARPES $E(k)$ intensity map with photon energy $h\nu$ = 130 eV along the path $\bar{\Gamma} - \bar{M} - \bar{X} - \bar{\Gamma}$ (inset, red) of the surface Brillouin zone. Band structure from DFT calculations for $k_z = 0$ is overlaid on top (dashed black) with bands linked to the decorated checkerboard model highlighted in green. The black arrow marks the weakly dispersing segment along X – Γ. **b** (left) ARPES $E(k)$ intensity map with $h\nu$ = 130 eV near $\bar{M}$ showing the $\alpha_M$ pocket and associated $2k_{F,\alpha}$. (right) DFT calculated band structure along the X – M − X path. **c** (left) ARPES $E(k)$ intensity map with $h\nu$ = 130 eV near $\bar{\Gamma}$ showing the $\beta_\Gamma$ pocket and associated $2k_{F,\beta}$. (right) DFT calculated band structure along the X – Γ − X path. **d** ARPES Fermi surface cross-section in the $k_x - k_y$ plane with $h\nu$ = 125 eV shows three closed pockets, labeled $\alpha_M$, $\beta_\Gamma$, and $\gamma_M$. Overlaid are circles (magenta) with diameters $2k_{F,\alpha}$ and $2k_{F,\beta}$ from the ARPES $E(k)$ maps, demarcating the $\alpha_M$ and $\beta_\Gamma$ pockets, respectively. **e** DFT calculated FS cross-sections for various $k_z$ projected into the $k_x - k_y$ plane. **f** Longitudinal resistance $\rho_{xx}(T)$ of bulk Pd$_5$AlI$_2$ showing metallic temperature dependence. (inset) Magnetoresistance $MR(H)$ at $T$ = 3 K showing prominent quantum oscillations above $\mu_0 H^\star \approx$ 3 T. **g** Fast Fourier transform $A_{FFT}$ of $\Delta MR(1/H)$ at $T$ = 3 K. (inset) Lifshitz-Kosevich fit (dashed line) to SdH amplitude versus $T$, $I_{SdH}(T)$ extracted at fixed $1/(\mu_0 H) = 0.16$ T$^{-1}$. **h** Hall effect $\rho_{yx}(H)$ up to $\mu_0 H$ = 36 T at $T$ = 0.5 K is characteristic of a multi-band system. We can identify signatures of the small, high-mobility $\alpha_M$ pocket at low $H$ (blue, shaded), and the large $\gamma_M$ electron pocket from the Hall slope at high $H$ (dashed line).

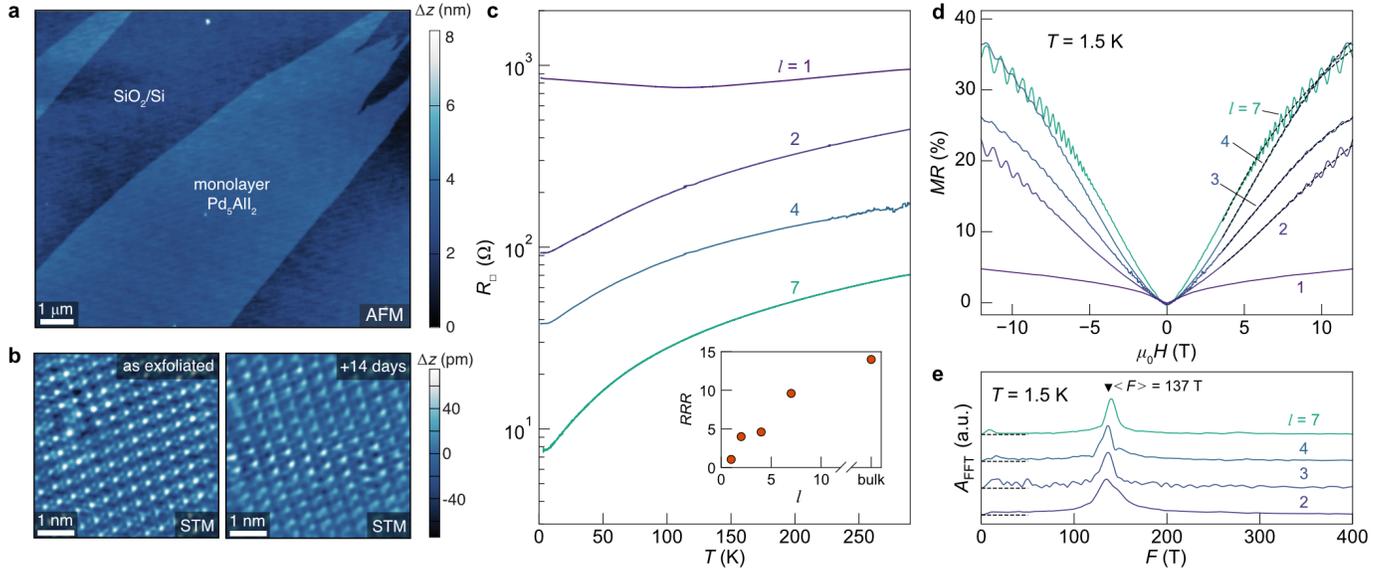

**Figure 3: Electronic transport and fermiology towards the monolayer limit**

**a** AFM image of Pd$_5$AlI$_2$ monolayer on SiO$_2$/Si wafer obtained by mechanical exfoliation in ambient conditions (Methods). **b** Atomic resolution STM images showing the square-lattice structure of Pd$_5$AlI$_2$ as exfoliated and after exposure to ambient conditions for 14 days. The latter image was obtained using a different STM tip. **c** Sheet resistance as a function of temperature $R_\square(T)$, evolving from strongly to weakly metallic with decreasing $l$; (inset) the residual resistivity ratio ($RRR$) as a function of $l$. **d** Magnetoresistance $MR(H)$ at $T = 1.5$ K for samples with $l$ spanning 1 to 7. A polynomial background (dashed black line) is subtracted to isolate the SdH oscillations. **e** Fast Fourier transforms $A_{FFT}$ of SdH oscillations observed in samples with $l$ spanning 2 to 7, vertically offset for clarity (dashed lines). All FFT spectra exhibit a prominent peak at an average frequency $\langle F \rangle \approx 137$ T, suggesting negligible variation of $E_F$ with $l$.

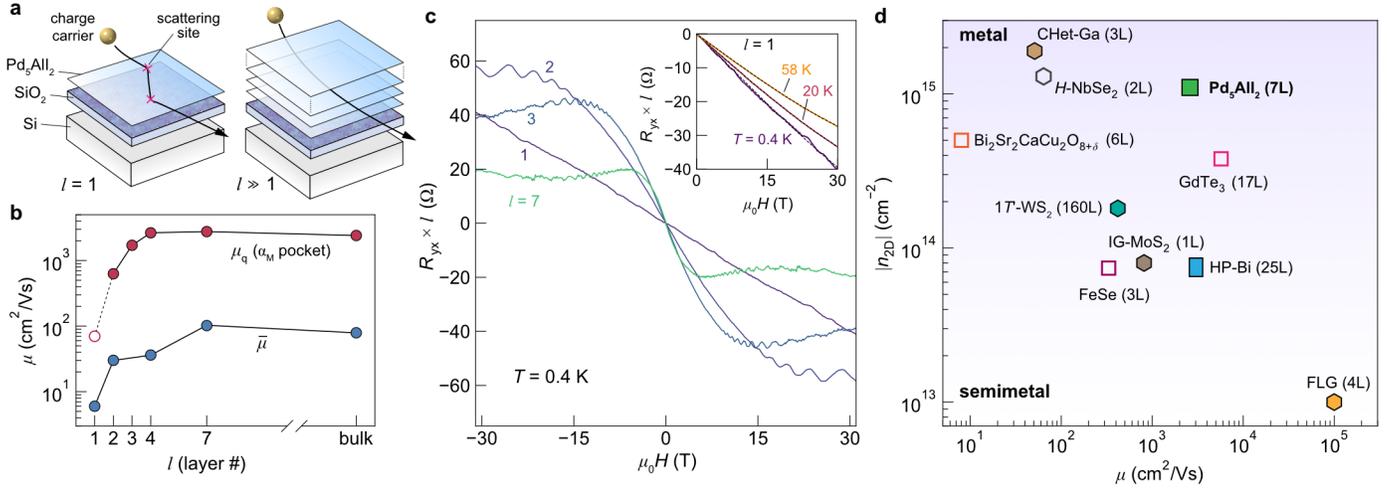

**Figure 4: Thickness-dependent Hall response and survey of vdW metals**

**a** Schematic of low $T$ transport in (left) monolayer and (right) multilayer devices. Electrons in the monolayer can encounter scattering sites (*e.g.,* adsorbates, surface roughness, etc.) that lead to localization, whereas in multilayer devices, parallel transport through interior layers can circumvent these sources of scattering. **b** Comparison of the average transport mobility $\bar{\mu}(l)$ and quantum mobility $\mu_q(l)$ of the $\alpha_M$ pocket (extracted from the $H$ dependent envelope of the oscillations, see Extended Data Fig. 8a) as a function of $l$. The open circle corresponds to the $\alpha_M$ pocket transport mobility obtained from a two-carrier Hall fit (see Fig. 4c, inset). **c** High-field Hall effect scaled by $l$, $R_{yx}(H) \times l$, at $T = 0.4$ K for devices with $l$ spanning 1 to 7, exhibiting an evolution from strong ($l = 7$) to weakly non-linear ($l = 1$) behavior. (inset) $R_{yx}(H)$ of device with $l = 1$ at different $T$ with two-carrier Hall fits shown as dashed lines (Methods). **d** Survey of vdW metals sorted by $\mu$ and 2D carrier density per layer $|n_{2D}|$ (see Extended Data Table 2). Number of layers in device is shown in parentheses. Open symbols are used for materials that are not air-stable when exfoliated.

**Methods**

**Synthesis of $Pd_5AlI_2$**

Palladium (Pd) powder (Fisher Scientific, 99.95%), aluminum (Al) chunks (Strem Chemicals, 99+%), and iodine ($I_2$) chunks (Aldrich, 99.99+%) were used as received for synthesis. In a typical reaction, 100 mg (0.940 mmol) Pd and 7 mg (0.259 mmol) Al were loaded into a 9 mm × 7 mm (O.D. × I.D.) fused silica tube in a nitrogen-filled glovebox. After removing the tube from the glovebox, 47.7 mg (0.188 mmol) $I_2$ was added after backfilling with argon (Ar). The tube was then sealed under ≈ 30 mTorr of Ar to a length of 15 cm. The portion of the tube containing starting materials was submerged in liquid $N_2$ while sealing to prevent loss of $I_2$. The sealed tube was placed horizontally in a box furnace, heated to 600 °C in 6 hrs, and held at this temperature for 3.5 days. Subsequently, the tube was water-quenched from 600 ºC. This reaction yielded shiny platelets of $Pd_5AlI_2$ with planar areas ranging from 1 – 100 $mm^2$.

**Scanning electron microscopy and energy dispersive X-ray spectroscopy**

Scanning electron microscope (SEM) images of single crystals were collected on a Zeiss Sigma VP SEM (Extended Data Fig. 2a). Energy dispersive X-ray spectroscopy (EDS) of single crystals was performed with a Bruker XFlash 6130 attachment. Spectra were collected with a beam energy of 15 keV. The surface of the crystal was freshly cleaved with Scotch tape before loading into the chamber. Elemental compositions and atomic percentages were estimated by integrating under the characteristic spectrum peaks for each element using Bruker ESPRIT 2 software (Extended Data Fig. 2b). Atomic percentages were calculated assuming only Pd, Al and I were present. Crystals were cleaved in air before data collection.

**Single crystal X-ray diffraction**

A single crystal of $Pd_5AlI_2$ was mounted onto MiTeGen Microloops LD with paratone oil. An Agilent Supernova single crystal diffractometer equipped with a Mo Kα micro-focus X-ray source and powered to 50 kV and 0.8 mA was used in the collection of diffraction data at $T$ = 101 K. The Crysalis–Pro software suite was used for optimizing data collection and performing integration and reduction of the data. ShelXT (50) and ShelXL (51) were used to solve and refine the crystal structure, respectively, with Olex2 (52) as the GUI to aid in solving the structure. The single crystal refinement results are summarized in Extended Data Table 1.

**Scanning tunneling microscopy**

Scanning tunneling microscopy (STM) measurements on freshly cleaved surfaces of bulk $Pd_5AlI_2$ crystals were conducted at $T$ = 292 K and in ultrahigh vacuum (base pressure below 2.0 × $10^{-10}$ Torr) using an Omicron VT-STM and electrochemically etched tungsten tips (annealed in vacuum before use). Topographic images were collected in constant current mode with bias voltage and current setpoints 200 mV and 50 pA, respectively. The cleaved crystal was removed from UHV conditions and stored in ambient lab conditions for 14 days. The sample was subsequently transferred back to the STM without additional cleaving or processing and scanned again.

**Electronic structure modelling**

Bulk density functional theory (DFT) calculations were performed using the Quantum Espresso (QE) software suite (53). We used ultrasoft pseudopotentials (54) and the Perdew–Burke–Erzenhof exchange-correlation functional (55). For the bulk calculations we used the structure provided by The Materials Project which is in agreement with the experimentally determined crystal and electronic structures. The calculation was performed on a $10 \times 10 \times 3$ $k$-mesh grid. We used publicly available tools to extract and process the electronic structure output from QE (56,57). Fermi surface cross-sections were obtained from DFT calculations using FermiSurfer (58).

The monolayer DFT electronic structure was calculated using the full-potential linearized augmented plane-wave (FP-LAPW) + local orbitals method implemented in the WIEN2k package (59) with the PBE exchange-correlation functional calculated within the generalized gradient approximation (PBE-GGA). The calculation was performed on a $37 \times 37 \times 7$ $k$-mesh grid. We set the plane-wave cut-off $K_{max}$ such that $R_{MT} \times K_{max} = 6.5$ where $R_{MT}$ is the smallest atomic sphere radius. We used $R_{MT}$ = 2.35 a.u., 2.5 a.u, and 2.5 a.u. for Al, Pd, and I respectively. The largest vector magnitude in the charge density Fourier expansion is set to $G_{max}$ = 12. The self-consistent calculation was run until the difference in charge density distribution between successive iterations fell below $10^{-5} \times e$ where $e$ is the elementary charge.

We now turn to the tight-binding modelling of the electronic structure. Broadly, two-dimensional bipartite lattices—lattices that can be partitioned into two sublattices such that there are only inter-sublattice connections—where the sublattices have an inequivalent number of sites can give rise to novel electronic structures from frustrated hopping (60). The prototypical models within this class are the two-dimensional (2D) dice (Extended Data Fig. 1a) and Lieb (Extended Data Fig. 1b) lattice models, where $t$ is the nearest-neighbor hopping energy. The frustrated hopping, specifically the absence of intra-sublattice bonds, together with the site imbalance between the sublattices, guarantees an electronic structure $E(k)$ with dispersive Dirac-like bands intersected at $E = 0$ by a dispersion-less flat band (Extended Data Figs. 1a and b).

The 2D decorated checkerboard presented in the main text (Fig. 1b) is also an imbalanced bipartite model with the central *d*-orbitals forming one sublattice and the corner *p*-orbitals forming another. Unlike the prototypical models, however, the frustrated intra-sublattice hopping of the decorated checkerboard arises from the orthogonality of the *d*-orbitals at the central site. Additionally, the rotational anisotropy of the *d*-orbitals causes a direction-dependent sign change of *t* and the appearance of two Dirac-like crossings at $E = 0$. Here, we describe the mapping between monolayer $Pd_5AlI_2$ and the decorated checkerboard model. In the absence of this orbital decoration, the decorated checkerboard reduces to a primitive square lattice (Extended Data Fig. 1c).

A notable aspect of flat bands in frustrated lattices is the fact that flat band eigenmodes are compact localized states (CLSs) created by quantum interference of electrons hopping on the lattice (1,61,62). Extended Data Figs. 1d – 1f shows the CLSs for the dice, Lieb, and decorated checkerboard model where interference of hopping leads to localization of electrons onto multi-site plaquettes. Indeed, this is a key distinction between flat bands arising from frustrated hopping and those in conventional heavy fermion compounds. In the latter, the flat band originates from electrons in highly localized in *f*-orbitals at a single site of the lattice.

We use a symmetry-based approach to construct a tight-binding model for the electronic structure of monolayer $Pd_5AlI_2$. Specifically, we focus on a minimal three-band model (excluding spin) which captures the pseudospin $S = 1$ Dirac structure observed at the M point in ARPES and DFT calculations (Figs. 2h and 2i). Given the 4/*mmm* ($D_{4h}$) point group symmetry of monolayer $Pd_5AlI_2$, we use a basis consisting of orbitals from the two-fold degenerate $E_g$ and singly degenerate $A_{2u}$ representations; the $E_g$ and $A_{2u}$ representations transforms like $[d_{xz}, d_{yz}]$ and $p_z$ orbitals, respectively, suggesting the former originates from the Pd atoms and the latter from the Al atoms. Due to the mirror symmetric structure across the *ab*-plane, these orbitals should be centered on the PdAl checkerboard lattice (Fig. 1g), resulting in the decorated checkerboard model (Fig. 1c). Extended Data Fig. 3a shows the resulting decorated checkerboard model including direction- and orbital-dependent sign changes of the hopping energy *t*. The band representations (63) of these orbitals reproduce the high-symmetry momentum representations obtain in the *ab-initio* calculations shown in Extended Data Fig. 3.

The nearest-neighbor (NN) Hamiltonian for spin-up electrons, with on-site atomic spin-orbit coupling (SOC) considered for the heavy Pd atoms, takes the form

$$\widehat{H}_0^\uparrow = \begin{array}{c|ccc} & d_{xz} & p_z & d_{yz} \\ \hline d_{xz} & 0 & 4t\,i\sin\left(\dfrac{k_x a}{2}\right)\cos\left(\dfrac{k_y a}{2}\right) & i\dfrac{\lambda}{2} \\ p_z & -4t\,i\sin\left(\dfrac{k_x a}{2}\right)\cos\left(\dfrac{k_y a}{2}\right) & 0 & -4t\,i\cos\left(\dfrac{k_x a}{2}\right)\sin\left(\dfrac{k_y a}{2}\right) \\ d_{yz} & -i\dfrac{\lambda}{2} & 4t\,i\cos\left(\dfrac{k_x a}{2}\right)\sin\left(\dfrac{k_y a}{2}\right) & 0 \end{array} \quad (S1)$$

where $t$ is the NN overlap integral between [$d_{xz}$, $d_{yz}$] and $p_z$ orbitals, $k_x$ and $k_y$ label in-plane crystal momentum, and $a$ is the $ab$-plane lattice constant; the spin-down Hamiltonian $\widehat{H}_0^\downarrow$ is the Hermitian conjugate of $\widehat{H}_0^\uparrow$. This model gives rise to an isolated perfectly flat band (Fig. 1c), even when $\lambda \neq 0$. Switching to the basis spanned by $d_\pm = -i/\sqrt{2}\left(d_{yz} \pm i d_{xz}\right)$ and $p_z$ followed by expanding Eq. S1 to linear order in $k$ around the M point $\left(k_x, k_y\right) = \left(\dfrac{\pi}{a}, \dfrac{\pi}{a}\right)$, we find the Hamiltonian (focusing on spin-up)

$$\widehat{H}_M^\uparrow = \begin{array}{c|ccc} & d_+ & p_z & d_- \\ \hline d_+ & \lambda/2 & \sqrt{2}ta(\kappa_x - i\kappa_y) & 0 \\ p_z & \sqrt{2}ta(\kappa_x + i\kappa_y) & 0 & \sqrt{2}ta(\kappa_x - i\kappa_y) \\ d_- & 0 & \sqrt{2}ta(\kappa_x + i\kappa_y) & -\lambda/2 \end{array} \quad (S2)$$

where ($\kappa_x$, $\kappa_y$) is in-plane crystal momentum measured from the M point, which can be rewritten in the form of pseudospin $S = 1$ massive Dirac Hamiltonian $\widehat{H} = \hbar v_F \, \mathbf{S} \cdot \boldsymbol{\kappa} + (\lambda/2) S_z$ (Eq. 1 of the main text); $\boldsymbol{\kappa} = <\kappa_x, \kappa_y>$, $v_F = 2ta/\hbar$ is Fermi velocity, and $\mathbf{S} = <S_x, S_y>$ with the $S = 1$ spin matrices given by

$$S_x = \dfrac{1}{\sqrt{2}}\begin{pmatrix} 0 & 1 & 0 \\ 1 & 0 & 1 \\ 0 & 1 & 0 \end{pmatrix}; \; S_y = \dfrac{1}{\sqrt{2}}\begin{pmatrix} 0 & -i & 0 \\ i & 0 & -i \\ 0 & i & 0 \end{pmatrix}; \; S_z = \begin{pmatrix} 1 & 0 & 0 \\ 0 & 0 & 0 \\ 0 & 0 & -1 \end{pmatrix} \quad (S3)$$

The solutions to this Hamiltonian are $E = 0$; $E = \pm \hbar v_F |\boldsymbol{\kappa}|$; there is a flat band separated from two dispersive branches by a gap $\Delta = \lambda/2$ in the presence of SOC. Expanding Eq. S1 around the $\Gamma$ point results in the same $S = 1$ Dirac Hamiltonian.

With the vector $\hbar v_F <\kappa_x, \kappa_y, \Delta/\hbar v_F> = |h| <\cos(\varphi)\sin(\theta), \sin(\varphi)\sin(\theta), \cos(\theta)>$ where $|h| = \sqrt{(\hbar v_F)^2(\kappa_x^2 + \kappa_y^2) + \Delta^2}$, $\tan(\varphi) = \kappa_y/\kappa_x$, and $\tan(\theta) = \Delta/\left(\hbar v_F \sqrt{\kappa_x^2 + \kappa_y^2}\right)$, the flat band eigenstate takes the form

$$|0\rangle = \frac{1}{|\kappa|} \begin{bmatrix} -e^{-i\varphi} \\ |\kappa| \tan(\theta) \\ e^{i\varphi} \end{bmatrix} \tag{S4}$$

The fact the flat band eigenstate is formed through interference between distinct orbitals, is an indication that it is a compact localized state (61). The positive (+) and negative (-) linearly dispersing eigenstates take the form

$$|\pm\rangle = \begin{bmatrix} \cos^2(\theta/2) e^{\mp i\varphi} \\ \sin(\theta)/\sqrt{2} \\ \sin^2(\theta/2) e^{\pm i\varphi} \end{bmatrix} \tag{S5}$$

Perturbations of Eq. S1 are needed to capture the electronic structure away from the high symmetry points. In particular, we consider next-nearest neighbor (NNN) hopping (Extended Data Fig. 3a) and distinct on-site energy of the $[d_{xz}, d_{yz}]$ and $p_z$ orbitals, $V_d$ and $V_p$, respectively, which make the flat band dispersive and shift the bands away from $E = 0$; their contribution to the spin-up sector, $\widehat{H}_1^\uparrow$, takes the form

$$\widehat{H}_1^\uparrow = \begin{bmatrix} V_d + 2t_1 \cos(k_x a) & 0 & 0 \\ 0 & V_p + 2t_2 \{\cos(k_x a) + \cos(k_y a)\} & 0 \\ 0 & 0 & V_d + 2t_1 \cos(k_y a) \end{bmatrix} \tag{S6}$$

where $t_1$ and $t_2$ are the NNN hopping energies among the $d$- and $p$-orbitals, respectively; the NNN contribution to the spin-down sector $\widehat{H}_1^\downarrow$ is identical since $\widehat{H}_1^\uparrow$ is diagonal. Extended Data Fig. 3b shows the band structure obtained from the combination of Eq. S1 and Eq. S6 using the parameters $t = 0.28$, $t_1 = 0.20$, $t_2 = 0.02$, $V_d = 0.30$, $V_p = 0.05$, and $\lambda = 0.05$. Examining the ARPES $E(k)$ intensity map and DFT results without SOC, we can trace bands near $E_F$ that are in qualitative agreement with the tight-binding (main text Fig. 2a, highlighted in green), indicating this tight-binding model captures the electronic structure around $E_F$. While the flat band becomes dispersive here due to higher-order hopping, it is noteworthy that a flat segment is preserved along the X – Γ line of the BZ (Extended Data Fig. 3b, arrow), with a corresponding peak in the density of states (Extended Data Fig. 3b). This suggests that, despite deviations from an ideal flat band, electron doping into this density of states peak could favor the emergence of novel correlated phases.

Similar agreement is seen between the measured and tight-binding Fermi surfaces, exhibiting three closed pockets (Extended data Fig. 3c). The relevance of the decorated checkerboard model to $Pd_5AlI_2$ is further supported by projecting the DFT bands onto the Pd $[d_{xz}, d_{yz}]$ (Extended Data Fig. 3d) and Al $p_z$ orbital (Extended Data Fig. 3e) of the PdAl checkerboard which reveals there is a large contribution from these orbitals to the bands at $E_F$. Moreover, examining thickness dependent electronic structure calculations (Extended Data Figs. 3f – 3i) from monolayer up to 4-layers (slab calculations without SOC), we find the overall electronic structure nearly invariant. This demonstrates the relevance of the decorated checkerboard model from the bulk to monolayer limit. Indeed, this is consistent with the SdH oscillations results shown in main text Fig. 3e where the oscillation frequency is essentially invariant versus $Pd_5AlI_2$ flake thickness. This is further supported by additional ARPES data (Extended Data Fig. 4) and angle-dependent quantum oscillations measurements (Extended Data Figs. 5f and 5g).

We now return to the decorated checkerboard model CLS (Extended Data Fig. 1f), and write down its explicit form within this tight-binding model. With the Pd atom of the bottom left unit cell as the origin, we can write this eigenstate as

$$A^\dagger = \frac{1}{\sqrt{8}} \begin{pmatrix} d_{xz}^\dagger(\vec{r}) + d_{yz}^\dagger(\vec{r}) \\ + d_{xz}^\dagger(\vec{r}+\vec{e}_x) - d_{yz}^\dagger(\vec{r}+\vec{e}_x) \\ - d_{xz}^\dagger(\vec{r}+\vec{e}_y) + d_{yz}^\dagger(\vec{r}+\vec{e}_y) \\ - d_{xz}^\dagger(\vec{r}+\vec{e}_x+\vec{e}_y) - d_{yz}^\dagger(\vec{r}+\vec{e}_x+\vec{e}_y) \end{pmatrix} \tag{S7}$$

where $\vec{e}_x = [a, 0]$ and $\vec{e}_y = [0, a]$ are the real-space lattice vectors while $d_{xz}^\dagger$ and $d_{yz}^\dagger$ are creation operators for $d_{xz}$ and $d_{yz}$ orbitals. Writing the decorated checkerboard Hamiltonian in real-space,

$$\widehat{H}_{\vec{r}} = t \begin{pmatrix} d_{xz}^\dagger(\vec{r}) p_z + d_{yz}^\dagger(\vec{r}) p_z \\ - d_{xz}^\dagger(\vec{r}+\vec{e}_x) p_z + d_{yz}^\dagger(\vec{r}+\vec{e}_x) p_z \\ + d_{xz}^\dagger(\vec{r}+\vec{e}_y) p_z - d_{yz}^\dagger(\vec{r}+\vec{e}_y) p_z \\ - d_{xz}^\dagger(\vec{r}+\vec{e}_x+\vec{e}_y) p_z - d_{yz}^\dagger(\vec{r}+\vec{e}_x+\vec{e}_y) p_z \end{pmatrix} + h.c. \tag{S8}$$

where $p_z$ is the annihilation operator for the $p_z(\vec{r} + 1/2(\vec{e}_x + \vec{e}_y))$ orbital, one can check that indeed $\widehat{H}_{\vec{r}}(A^\dagger | vac.\rangle) = 0$ ($| vac.\rangle$ is the vacuum state). The CLS can be translated to any unit cell in the crystal without energy cost, indicating there is a macroscopically degenerate manifold of states with $E = 0$.

Turning to electrical transport, the low $T$ behavior in the monolayer limit indicates that the $\alpha_M$ and $\beta_\Gamma$ pockets from the Dirac-like branches of $S = 1$ DFs remain mobile, while electrons of the large $\gamma_M$ pocket are localized. This raises the question whether electronic mobility of the $\alpha_M$ and $\beta_\Gamma$ pocket $S = 1$ DFs may be protected against backscattering, similar to $S = \frac{1}{2}$ Dirac fermions in graphene, relative to the $\gamma_M$ electrons.

Broadly, disorder scattering during the transport process transfers a carrier with momentum $k$ on the Fermi surface to another state with momentum $k'$. The dominant effect on electrical transport is backscattering (*i.e.* large-angle scattering) where $k' = -k$. Correspondingly, if the wavefunction overlap between states at $k$ and $-k$ is minimized, backscattering and the influence of disorder on transport is suppressed.

Massless $S = ½$ Dirac fermions like in graphene enjoy fully suppressed backscattering due to the orthogonality of eigenstates at $k$ and $-k$ (64,65). Further, when a mass gap is opened in graphene, $k$ and $-k$ eigenstates are no longer orthogonal and protection against backscattering is weakened. For the linearly dispersing branches of $S = 1$ DFs (Eq. S5), corresponding to the $\alpha_M$ and $\beta_\Gamma$ pockets, the wavefunction overlap of momentum-reversed states takes the form

$$\langle \pm_{-k} | \pm_k \rangle = \frac{1}{2}\left(1 + \frac{\Delta^2}{E_F^2}\right) \tag{S9}$$

Unlike $S = ½$ Dirac fermions, $S = 1$ DFs are partially protected against backscattering with a minimum 50% overlap between momentum-reversed eigenstates. For $\Delta \neq 0$, the overlap increases and protection is weakened (Extended Data Fig. 3j). For the $\alpha_M$ pocket, we estimate from ARPES $\Delta_\alpha \approx 80$ meV and $\Delta_\alpha/E_F \approx 0.1$ (main text Fig. 2b), indicating they experience almost the maximum possible protection against backscattering for $S = 1$ DFs. For the $\beta_\Gamma$ pockets, however, the corresponding gap estimated from DFT calculations (main text Fig. 2c) appears to be larger, $\Delta_\beta \approx 240$ meV. From the $E(k)$ map near $\bar{\Gamma}$ (main text Fig. 2c), we extract $E_F = \hbar v_F k_F \approx 285$ meV given $v_{F,\beta} \approx 3.1 \times 10^5$ and $k_{F,\beta} \approx 0.14$ Å$^{-1}$. With $\Delta_\beta/E_F \approx 0.84$, the $\beta_\Gamma$ pocket $S = 1$ DFs experiences partial protection against backscattering, but less than the $\alpha_M$ pocket.

Nevertheless, the partial protection of the $\alpha_M$ and $\beta_\Gamma$ pockets is in stark contrast to conventional dispersive bands like that forming the $\gamma_M$ pocket, which do not enjoy any protection (Extended Data Fig. 3j). This suggests disorder-induced backscattering affects the conventional $\gamma_M$ electronics more strongly than $S = 1$ DFs of the $\alpha_M$ and $\beta_\Gamma$ pockets, potentially causing localization of the former while the latter remain mobile.

**Bulk ARPES**

ARPES measurements were performed at the Electron Spectro-Microscopy (ESM) 21-ID-1 beamline of the National Synchrotron Light Source II, USA. Bulk crystals of Pd$_5$AlI$_2$ crystals were cleaved inside an ultra-high vacuum chamber using Kapton tape before the ARPES experiments. The beamline is equipped with a Scienta DA30 electron analyzer, with base pressure $\sim 2 \times 10^{-11}$ mbar. The total energy resolution for ARPES measurements was approximately 12 meV and the angular resolution was better than 0.1° and 0.3° parallel and perpendicular to the slit of the analyzer, respectively. All the measurements were at $T \approx$ 15 K in linear horizontal plus linear vertical polarized modes (LH + LV) unless otherwise stated.

Extended Data Figs. 4a – 4c shows $E(k)$ measured by ARPES for photon energies $hv$ = 127 eV to 142 eV, which probe distinct ranges of out-of-plane momentum $k_z$, demarcated in Extended Data Fig. 4e as dashed lines; we assume an inner potential of 10 eV to estimate $k_z$. The weak variation of these $E(k)$ spectra is indicative of a quasi-2D electronic structure. This is further supported by examining the Fermi surface cross-sections in the $k_x$ – $k_y$ plane at various $k_z$, controlled by $hv$ (Extended Data Fig. 4d, measured with LV polarization). Indeed, the quasi-2D nature is even more evident in a $k_\parallel$ – $k_z$ map of the FSs along the $\overline{M}$ – $\overline{\Gamma}$ – $\overline{M}$ line (Extended Data Fig. 4e), where we observe weak variation along the entire $k_z$ span of the BZ, characteristic of quasi-2D, cylindrical Fermi surfaces. We also examined the electronic structure away from $E_F$ by ARPES and DFT calculations. Extended Data Figs. 4f – 4h show Fermi surface cross-sections in the $k_x$ – $k_y$ plane, across a 750 meV span of $E$ below $E_F$, together with comparison to DFT results, Extended Data Figs. 4i – 4k. Similar to the features at $E_F$, we observe satisfactory agreement between theory and experiment.

**Bulk magnetotransport and magnetization measurements**

$Pd_5AlI_2$ single crystals were mounted on glass coverslips using GE varnish. Electrical contact was made using 25 μm diameter gold wire (99.99% purity) and silver paint (Dupont 4929N) in a standard Hall bar geometry. In-house transport measurements were performed in a commercial cryostat equipped with a superconducting magnet (Quantum Design Dynacool) using the Keithley 6221 current source and 2182A nanovoltmeter delta mode technique. Longitudinal ($\rho_{xx}$) and Hall ($\rho_{yx}$) resistivity were symmetrized and anti-symmetrized, respectively, to eliminate the effects of contact misalignment. High field transport measurements at $T$ = 0.5 K and up to $\mu_0 H$ = 36 T were performed in Cell 14 at the National High Magnetic Field Laboratory in Tallahassee, FL.

As discussed in the main text, the low $H$ quadratic magnetoresistance can be used to extract $\mu$ for the highest $\mu$ carriers given $MR(H) = \mu^2(\mu_0 H)^2$ (66). Extended Data Fig. 5a shows $MR(H)$ at $T$ = 3 K in the range $|\mu_0 H|$ ≤ 0.5 T along with a quadratic fit (dashed line) obtained by least-squares regression which yields $\mu$ = 3,460 cm$^2$/Vs which we associate with the $\alpha_M$ pocket carriers given they exhibit quantum oscillations.

Quantum oscillations appear as $1/H$ periodic variations in transport or thermodynamic observables of materials due to Landau quantization of the electronic structure (31). The oscillation frequency $F$ is linked to the cross-sectional area $A_F$ of the Fermi pocket generating the quantum oscillations through Onsager's relation $F = \hbar A_F/2\pi e$ where $\hbar$ is the reduced Planck's constant. We extract the SdH quantum oscillations by subtracting a polynomial background and perform a fast Fourier transform (FFT) to obtain the $F$ spectrum.

The temperature dependence of quantum oscillations can be used to extract the effective mass $m^\star$ through the conventional Lifshitz-Kosevich analysis. With increasing $T$, the amplitude of the $1/H$ periodic oscillations are damped by the Lifshitz-Kosevich factor which takes the form (31)

$$A_{LK}(T,H) = \frac{X}{\sinh(X)} \tag{S10}$$

where $X = \eta T/\mu_0 H \left(m^\star/m_e\right)$, with $m_0$ the bare electron mass and the constant $\eta$ = 14.69 T/K. Consequently, the effective mass $m^\star$ can be extracted by tracking the SdH quantum oscillation amplitude $I_{SdH}(T)$ at fixed $H$ and fitting to Eq. S10. Extended Data Fig. 5b shows SdH oscillations $\Delta MR(1/H)$ at various fixed $T$ isolated by subtracting a 3$^{rd}$ order polynomial background from $MR(H)$; extracting $I_{SdH}(T)$ at $\mu_0 H^\star$ = 3 T (see Fig. 2g, inset) and fitting to Eq. S10 yields $m^\star \approx 0.16\ m_0$.

While the $T$ dependence at fixed $H$ provides insight into the effective mass, the $H$ dependence at fixed $T$ can be used to extract the quantum mobility $\mu_q$ of the Fermi surface generating the oscillations. Specifically, the envelope of the oscillations exhibits an exponential suppression of the form (31)

$$\Delta MR\left(1/H\right) \propto e^{-\pi/\mu_q(\mu_0 H)} \tag{S11}$$

Extended Data Fig. 5c shows $\Delta MR(1/H)$ at $T$ = 3 K along with a least-squares fit of the envelope to Eq. S11, wherein we obtain $\mu_q$ = 2,411 cm$^2$/Vs. The fact that the quantum mobility is less than the transport mobility extracted from $MR(H)$ at the low $H$ is consistent with the fact that the transport mobility is only sensitive to large $k$ scattering (42).

DC magnetization $M$ was measured in a Quantum Design PPMS Dynacool system equipped with a vibrating sample magnetometer. Two single crystals of Pd$_5$AlI$_2$ were mounted on a quartz platform and attached to a diamagnetic brass sample holder using GE varnish with the $c$-axis oriented parallel to $H$. $M(H)$ was measured at $T$ = 2 K up to $\mu_0 H$ = 9 T (Extended Data Fig. 5d) and a diamagnetic correction was applied to account for contributions from the sample holder. De Haas-van Alphen (dHvA) oscillations $\Delta M(1/H)$ (Extended Data Fig. 5d) were isolated from $M(H)$ by subtracting a polynomial background; the dHvA oscillation frequency $F \approx 130$ T was extracted by performing an FFT of $\Delta M(1/H)$ (Extended Data Fig. 5d, inset). The variance in $F$ between bulk SdH, bulk dHvA, and nanodevice SdH oscillations suggests there is batch-to-batch variation of $E_F$. From the measured $F$'s across these samples and $v_F$ obtained from DFT, we estimate $E_F$ varies by ±2.5 %.

While we observe oscillations from the small α$_M$ pocket of Pd$_5$AlI$_2$, scrutinizing $MR(H)$ and $\rho_{yx}(H)$ to higher $H$ up to 36 T (Extended Data Fig. 5e) does not reveal additional quantum oscillations from the β$_\Gamma$ and γ$_M$

Fermi surfaces. This suggests the quantum mobilities of these two pockets are less than $1/36$ $T^{-1} \approx 280$ cm$^2$/Vs. It is also possible that the amplitude of these oscillations are weaker relative to the background $MR(H)$ and $\rho_{yx}(H)$, making them challenging to observe. Measuring up in pulsed magnetic field up to ~ 100 T could help reveal these oscillations.

Extended Data Fig. 5f shows the FFT spectra obtained from SdH oscillations measured at $T = 0.4$ K and various fixed angles $\theta$ measured relative to the $c$-axis (Extended Data Fig. 5f, inset), plotted versus $F\cos(\theta)$. The fact that the FFT spectra are aligned when plotted on this scaled axis indicates the oscillations stem from a cylindrical Fermi surface. Indeed, we can fit the measured $F(\theta)$ to the expected $1/\cos(\theta)$ dependence (Extended Data Fig. 5g) (31).

The SdH oscillations survive up to $\theta \sim 56°$ within the measured range $|\mu_0 H| \leq 36$ T, suggesting measurements up to higher $H$ could be used to further map $F(\theta)$. However, this data together with the ARPES Fermi surface cross-section establishes the cylindrical, quasi-2D nature of the bulk electronic structure. Indeed, previously reported interlayer transport measurements show that the out-of-plane resistivity is five orders of magnitude larger than the in-plane resistivity, further evidencing quasi-2D transport behavior (22).

It is also noteworthy that, examining the ARPES FS cross-sections (main text Fig. 2d), the $\gamma_M$ pocket and the $\beta_\Gamma$ pocket exhibit relatively small momentum space separation. Schematically, this would give rise to open-orbits parallel to the $\Gamma - X$ direction, evidenced by non-saturating magnetoresistance (66) and the abrupt emergence of new quantum oscillation frequencies at large $H$ (67,68); we do not observe these characteristic features in the measured $H$ range. It may be interesting future work to push to higher field using pulsed techniques to induce magnetic breakdown in this robust 2D metal, complementing quasi-2D organics (69).

**Exfoliation, AFM characterization, and thickness calibration**

Flakes of Pd$_5$AlI$_2$ were mechanically exfoliated using Scotch tape which was subsequently pressed onto 285 nm SiO$_2$/Si substrates pre-etched in O$_2$ plasma for 5 mins. After baking on a hot plate for 1 min at 100 ºC, the tape was peeled off slowly from the substrate, leaving behind flakes as thin as monolayer Pd$_5$AlI$_2$. We initially determined flake thickness using AFM (Bruker Dimension FastScan operated in tapping mode); Extended Data Fig. 6a shows an AFM scan of a terraced Pd$_5$AlI$_2$ with sharp edges across which thickness changes by $\approx 1.3$ nm (Extended Data Fig. 6b); we associate these to single Pd$_5$AlI$_2$ layer steps, in nominal agreement with the layer thickness $\approx 0.97$ nm determined from single crystal X-ray diffraction (1/2 the $c$-axis lattice constant).

Few layer flakes exfoliated onto 285 nm $SiO_2$/Si substrates exhibit optical contrast in microscope images (Extended Data Fig. 6c) $C_l^i = I_l^i - I_0^i / I_0^i$, where $I_l^i$ is the intensity for a $l$ layer flake in the $i^{th}$ color channel ($i$ = red, blue, or green). Extended Data Fig. 6d shows $C_l^i$ averaged in regions of distinct thickness (Extended Data Fig. 6c, labelled squares) as identified by AFM topography with error bars corresponding to the standard deviation. Although the red channel cannot be used differentiate flakes with $l = 1 - 3$, the blue and green contrast are sufficiently distinct to identify monolayers; for $l \geq 2$ the contrast tends to saturate which makes differentiation more challenging. However, after spin-coating 500 nm of PMMA for device fabrication (Extended Data Fig. 6e), the optical contrast is enhanced in all three channels due to the additional PMMA-flake interface; Extended Data Fig. 6f shows $C_l^i$ averaged in the same regions, which is clearly distinct up to $l = 3$.

**Device fabrication and transport**

Few layer flakes of $Pd_5AlI_2$ were exfoliated onto 285 nm $SiO_2$/Si substrates, pre-treated for 5 mins in $N_2/O_2$ plasma, using standard Scotch tape techniques. The substrates were spin coated with 500 nm of PMMA resist and baked for 3 mins at 160 ºC. After identifying flake thicknesses using optical contrast, device patterns were written using electron beam lithography and developed in cold IPA/$H_2O$. Electrical contacts were made by electron beam evaporation of Cr/Pd/Au (typically 5 nm/15 nm/70 nm) followed by lift-off in acetone. We found that $Pd_5AlI_2$ flakes exfoliated onto silinated $SiO_2$/Si substrates (70) can be manipulated using standard vdW dry-transfer techniques (71) to construct heterostructures, see for example Extended Data Fig. 8d.

Transport measurements down to $T = 1.6$ K were performed in a commercial cryostat equipped with a superconducting magnet (Oxford Instruments). Measurements down to $T = 0.4$ K (0.3 K) and up to $\mu_0 H = 31$ T ($\mu_0 H = 36$ T) were performed in Cell 9 (Cell 14) at the National High Magnetic Field Laboratory in Tallahassee, FL.

As noted in the main text, a notable aspect of $R_\square(T)$ in the monolayer $l = 1$ device (main text Fig. 3c) is the increase in $R_\square$ for decreasing $T$ below approximately 100 K. The change in $R_\square$ below 100 K is relatively small, increasing by 13% from 100 K to 1.6 K, which is inconsistent with the exponential increase anticipated if there is a bandgap near $E_F$. Instead, examining the sheet conductance $G_\square = 1/R_\square$ versus $\ln(T)$ (Extended Data Fig. 7a), we find a linear regime just below 100 K which flattens into a plateau-like regime below 13 K. The regime with $G_\square \propto \ln(T)$ is consistent with charge carrier localization by disorder (72) and resembles the behavior of evaporated metallic films (73). Furthermore, the mobility analysis (main text Fig. 4b) indicates the $\gamma_M$ pocket electrons are the ones being localized. We attribute the plateau-like feature below 13 K to shunting of electrical transport by the $\alpha_M$ and $\beta_\Gamma$ pocket $S = 1$ DFs which remain mobile at this $T$ (main text Fig. 4b). The weak decrease in $G_\square$ below 13 K may signal incipient localization of the $\beta_\Gamma$ pocket which have lower $\mu$ and are less protected against disorder scattering relative to the $\alpha_M$ carriers.

Evidence of weak localization is also observed in the magnetotransport response. Examining the magnetoconductance $\Delta\sigma(H) = \Delta\sigma(H) = 1/R_\square(H) - 1/R_\square(0)$ (valid since Pd$_5$AlI$_2$ is in the small Hall angle regime) for $l = 1$, for $|\mu_0 H| \leq 9$ T and at various fixed $T$ (Extended Data Fig. 7b) we see a pronounced cusp at low $H$ developing at $T \lesssim 20$ K. This behavior is well-described by the Hikami-Larkin-Nagaosa theory for weak anti-localization (74) in metals where $\Delta\sigma(H)$ is given by

$$\Delta\sigma(H) = \frac{\alpha e^2}{\pi\hbar}\left[\psi\left(\frac{l_B^2}{l_\varphi^2} + \frac{1}{2}\right) - \ln\left(\frac{l_B^2}{l_\varphi^2}\right)\right] \tag{S12}$$

with the magnetic length $l_B^2 = \hbar/eB$, $l_\varphi$ is the phase coherence length, and $\alpha$ is a coefficient on the order of unity; fits to Eq. S12 at each $T$, obtained by least-square regression, are shown as dashed lines in Extended Data Fig. 7b. Similar behavior has been seen in thin vdW metals such as Bi$_2$Se$_3$ (75) and VSe$_2$ (76) however, these systems are not air-stable. For thicker devices with $l = 2$ to 7, the low $H$ magnetoresistance is instead quadratic following $MR(H) = \mu_\alpha^2(\mu_0 H)$ as seen in Extended Data Fig. 7c, $MR(H)$ at $T = 3$ K in the range $|\mu_0 H| \leq 0.5$ T, from which we estimate $\mu_\alpha(l)$ by fitting to (dashed lines) by least-squares regression; $\mu_\alpha(l)$ extracted from these fits is shown in the inset of Extended Data Fig. 7c, complementing $\mu_q(l)$ of the $\alpha_M$ pocket shown in the inset of Fig. 3b and discussed further below; $\mu_q(l) < \mu_\alpha(l)$ from low-$H$ quadratic $MR$ likely reflects the fact that $\mu_q$ is sensitive to scattering channels that cause dephasing unlike $\mu_\alpha$.

We now turn to SdH oscillations $\Delta MR(1/H)$ from devices with $l = 2$ to 7. After isolating $\Delta MR(1/H)$ from $MR(H)$ using a 3$^{rd}$ order polynomial background for various $l$ at $T = 1.5$ K (Extended Data Fig. 8a). We estimate the quantum mobility $\mu_q(l)$ of the $\alpha_M$ pocket (main text Fig. 4b) by least-squares fitting the exponential suppression of oscillation amplitude with $1/H$ to the anticipated exponential suppression Eq. S11, the results of which are shown as dashed lines in Extended Data Fig. 8a.

We measure the $T$-dependence of $\Delta MR(1/H)$ for $l = 2$ to evaluate $m^\star$ in the few-layer limit. Extended Data Fig. 8b shows $\Delta MR(1/H)$ for a device with $l = 2$ at fixed $T$ spanning 0.4 K to 60 K. We extract the oscillation amplitude $I_{SdH}(T)$ at fixed $(\mu_0 H)^{-1} = 0.049$ T$^{-1}$ (Extended Data Fig. 8b, black triangle) and perform a least-square fit to the Lifshitz-Kosevich form Eq. S10 (Extended Data Fig. 8c). The fit yields $m^\star = 0.18\ m_0$ which is in good agreement with bulk samples and further demonstrates the electronic structure does not vary as the layer number is reduced.

The gate dependence of quantum oscillations in exfoliated flakes further demonstrates that the electronic structure is consistent with bulk Pd$_5$AlI$_2$. Extended Data Fig. 8d shows a 3-layer Pd$_5$AlI$_2$ transport device, where a backgating geometry is used to control the carrier density during transport measurements. Examining $MR(H)$ up to high-field $\mu_0 H = 36$ T, we find that SdH oscillations shift gradually with backgate voltage $V_{BG}$ (Extended Data Fig. 8e). Specifically, an FFT analysis of the oscillations (Extended Data Fig. 8e, inset) reveals the oscillation frequency increases with increasing $V_{BG}$. A positive, linear relationship between $V_{BG}$ and the oscillation $F$ is clearly observed across the measured $V_{BG}$ values, consistent with electron-like carriers, as expected for the $\alpha_M$ pocket.

**Carrier densities and multi-band Hall analysis**

As noted in the main text, the Hall slope $d\rho_{yx}/dH$ at high $H$ (main text Fig. 2h) is linked to the difference in total carrier density between electrons ($n_e$) and holes ($p_h$), $d\rho_{yx}/dH = 1/(e(n_e - p_h))$ where $e$ is the elementary charge. Correspondingly, from the linear fit to $\rho_{yx}(H > 20$ T) (main text Fig. 2h, dashed line) we extract $(n_e - p_h) \approx 8.7 \times 10^{21}$ cm$^{-3}$. Using the relation

$$|n_{2D}| = |n_{3D}| \times t \tag{S13}$$

where $t \approx 1$ nm is the thickness of monolayer Pd$_5$AlI$_2$, we obtain the total 2D density difference per layer $(n_e - p_h)_{,2D} = 9.4 \times 10^{14}$ cm$^{-2}$. Conversion from 3D to 2D using Eq. S13 is permitted by the fact that the electronic structure is quasi-2D (Extended Data Fig. 4e).

This result for the density difference can be combined with the ARPES $k_x - k_y$ map (Fig. 2d) to estimate the carrier density contributed by each pocket. In particular, Luttinger's theorem states the Fermi surface cross-sectional area $A_F$ is proportional to the 2D carrier density of the pocket, $|n_{2D}| = gA_F/4\pi^2$, where $g$ is the degeneracy of the pocket. From area ratios extracted from the ARPES $k_x - k_y$ map (main text Fig. 2d) we obtain the density ratios $|n^\alpha_{2D}|/|n^\gamma_{2D}| \sim 1 \times 10^{-2}$ and $|p^\beta_{2D}|/|n^\gamma_{2D}| \sim 0.11$. Correspondingly, the carrier densities must satisfy the equation

$$(n_e - p_h)_{2D} = n^\gamma_{2D} \times \left(1 - \frac{|p^\beta_{2D}|}{|n^\gamma_{2D}|} + \frac{|n^\alpha_{2D}|}{|n^\gamma_{2D}|}\right) \tag{S14}$$

From Eq. S14 we find $n^\alpha_{2D} \sim 9.7 \times 10^{12}$ cm$^{-2}$, $p^\beta_{2D} \sim 1.1 \times 10^{14}$ cm$^{-2}$ (holes), and $n^\gamma_{2D} \sim 9.7 \times 10^{14}$ cm$^{-2}$, respectively. The corresponding total 2D and 3D carrier densities are $|n_{2D}| \approx 1.1 \times 10^{15}$ cm$^{-2}$ and $|n_{3D}| \approx 1.1 \times 10^{22}$ cm$^{-3}$. Notably, $n^\alpha_{2D}$ can also be extracted using the measured $A_F$ of the $\alpha_M$ pocket from SdH oscillations: from $F = 140$ T and assuming a degeneracy $g = 2$ for spin, we find $n^\alpha_{2D} \sim 7 \times 10^{12}$ cm$^{-2}$ from SdH oscillations which is consistent with the value obtained from ARPES and Luttinger's theorem for quasi-2D systems.

We now turn to estimates of average mobility $\bar{\mu}$ in $Pd_5AlI_2$. For a single band system with carrier density $n$, the Drude relation $\rho = 1/\sigma = 1/ne\mu$ can be used to extract the mobility. In a multi-band system, the individual mobilities become challenging to extract, however, an average mobility $\bar{\mu}$ can be obtained from a similar analysis. For a multi-band system, the Drude relation takes the form

$$\rho = \frac{1}{\sigma} = \frac{1}{\sum_i n_i e \mu_i} = \frac{1}{n_{tot} e \bar{\mu}} \tag{S15}$$

where $n_i$ and $\mu_i$ are the density and mobility of the $i^{th}$ band, $n_{tot} = \sum_i n_i$ is the total carrier density, and $\bar{\mu}$ is an average mobility given by

$$\bar{\mu} = \frac{\sum_i n_i \mu_i}{n_{tot}} \tag{S16}$$

From Eq. S16 we can see that $\bar{\mu}$ is dominated by the band contributing the largest number of carriers, which is the $\gamma_M$ pocket in $Pd_5AlI_2$.

For devices with $l > 2$, we use SdH oscillations to gain insight into the mobility of the $\alpha_M$ pocket $S = 1$ DFs (Extended Data Fig. 8a). For the monolayer device, however, the absence of SdH oscillations requires the use of an alternative approach. As noted in the main text, non-linear $\rho_{yx}(H)$ is typical of multi-band systems with coexisting electron and hole FSs. In typical metals with small Hall angle (i.e., $\rho_{yx}/\rho_{xx} \ll 1$), $\rho_{yx}(H)$ takes the form (66)

$$\rho_{yx} = \frac{\sum_i -\delta_i \sigma_i \gamma_i / 1 + \gamma_i^2}{\left(\sum_i \sigma_i / 1 + \gamma_i^2\right)^2} \tag{S17}$$

with $\sigma_i = n_i e \mu_i$ ($p_i e \mu_i$), $\gamma_i = \mu_i \mu_0 H$, and $\delta_i = +1$ (−1) for holes (electrons). In general, it is challenge to directly fit measured $\rho_{yx}(H)$ to Eq. S17. In the present case, however, the constraints from ARPES and Luttinger's theorem on the carrier densities for each FS can assist in extracting $\mu_i$ using Eq. S17. Indeed, in the monolayer device where the mobility of the $\gamma_M$ pocket is small ($< 6$ cm$^2$/Vs as estimated from the average mobility, main text Fig. 4b) and $R_{yx}(H)$ weakly non-linear (main text Fig. 4c, inset), we can focus on the $\alpha_M$ and $\beta_\Gamma$ pockets and utilize a two-band model for $R_{yx}(H)$ (66) to estimate $\mu_\alpha$ and $\mu_\beta$,

$$R_{yx}(H) = -\left(\frac{\mu_0 H}{en_{2D}^{\alpha}}\right)\frac{\eta_{\alpha}^2 - c\eta_{\beta}^2 + (1-c)\eta_{\alpha}^2\eta_{\beta}^2}{(\eta_{\alpha} + c\eta_{\beta})^2 + (1-c)\eta_{\alpha}^2\eta_{\beta}^2} \tag{S17}$$

with $\eta_{\alpha} = \mu_{\alpha}(\mu_0 H)$, $\eta_{\beta} = \mu_{\beta}(\mu_0 H)$, and $c = \left(n_{2D}^{\beta}/n_{2D}^{\alpha}\right) \sim 11$ with $n_{2D}^{\alpha}$ fixed at $9.7 \times 10^{12}$ cm$^{-2}$ by the previous Hall analysis. The fits to Eq. S17 obtained by least-square regression are shown as dashed lines in main text Fig. 4c (inset). At $T = 0.4$ K, we estimate $\mu_{\alpha} \approx 70$ cm$^2$/Vs and $\mu_{\beta} \approx 15$ cm$^2$/Vs in the monolayer device. The fact that the $\beta_\Gamma$ pocket mobility is nearly five times smaller than that of the $\alpha_M$ pocket is consistent with the weak non-linearity of $R_{yx}(H)$, closely resembling a single carrier Hall effect.

**Survey of exfoliated vdW materials**

Extended Data Table 2 summarizes the survey of exfoliated vdW materials sorted by 2D carrier density per layer $|n_{2D}|$ and $\mu$ shown in Fig. 4d. Here we outline the methods used to construct this survey. **[Chet-Ga (3L)]** This is 3-layer elemental gallium grown under graphene layers on SiC (45). We estimate $|n_{2D}|$ using Fermi surface cross-sectional areas mapped by ARPES data in ref. (45) and Luttinger's theorem. Subsequently we estimate $\mu$ using the relation

$$\mu = \frac{1}{R_{xx} e |n_{2D}|} \tag{S18}$$

with $R_{xx}$ the longitudinal resistance provided in the same report. **[$H$-NbSe$_2$ (2L)]** Exfoliated transition metal dichalcogenide (TMD) vdW metal. We estimate $|n_{2D}|$ using Eq. S18 with $t \approx 0.65$ nm the approximate thickness of $H$-NbSe$_2$ layer with $|n_{3D}|$ reported in ref. (77). We extract $\mu$ using transport data in ref. (78) and Eq. S15. **[Bi$_2$Sr$_2$CaCu$_2$O$_{8+\delta}$ (6L)]** Layered high-temperature superconductor. $|n_{2D}|$ and $\mu$ are reported in ref. (79). **[FeSe (3L)]** Layered iron-based superconductor. We estimate $|n_{2D}|$ using Eq. S18 where $t \approx 0.59$ nm is the thickness of a single FeSe layer with $|n_{3D}|$ for a 3 layer device reported in ref. (80); we estimate $\mu$ using Eq. S18 from transport data in the same report. **[GdTe$_3$ (17L)]** Anti-ferromagnetic vdW metal. We estimate $|n_{2D}|$ using Eq. S18 with $t \approx 1.27$ nm the thickness of a single GdTe$_3$ layer and $|n_{3D}|$ for a 17L device reported in ref. (81); $\mu$ for this device is provided in the same report. **[1$T'$-WS$_2$ (160L)]** We estimate $|n_{2D}|$ using Eq. S18 with $t \approx 0.57$ nm is the thickness of a single 1$T'$-WS$_2$ layer with $|n_{3D}|$ for a 160 layer thick device reported in ref. (82). The $\mu$ for this device is provided in the same report. **[IG-MoS$_2$ (1L)]** Ionic-liquid gated TMD semiconductor $H$-MoS$_2$. We show the maximum $\mu$ and corresponding $|n_{2D}|$ reported in ref. (83). **[HP-Bi (25L)]** Hot-pressed elemental bismuth. We extract $\mu$ from the onset of quantum oscillations and $|n_{2D}|$ using Eq. S18 from transport data for a 7 nm thick flake reported in ref. (84). **[FLG (4L)]** Few-layer graphene exfoliated from graphite. $|n_{2D}|$ and $\mu$ reported in ref. (37).

**Extended references**

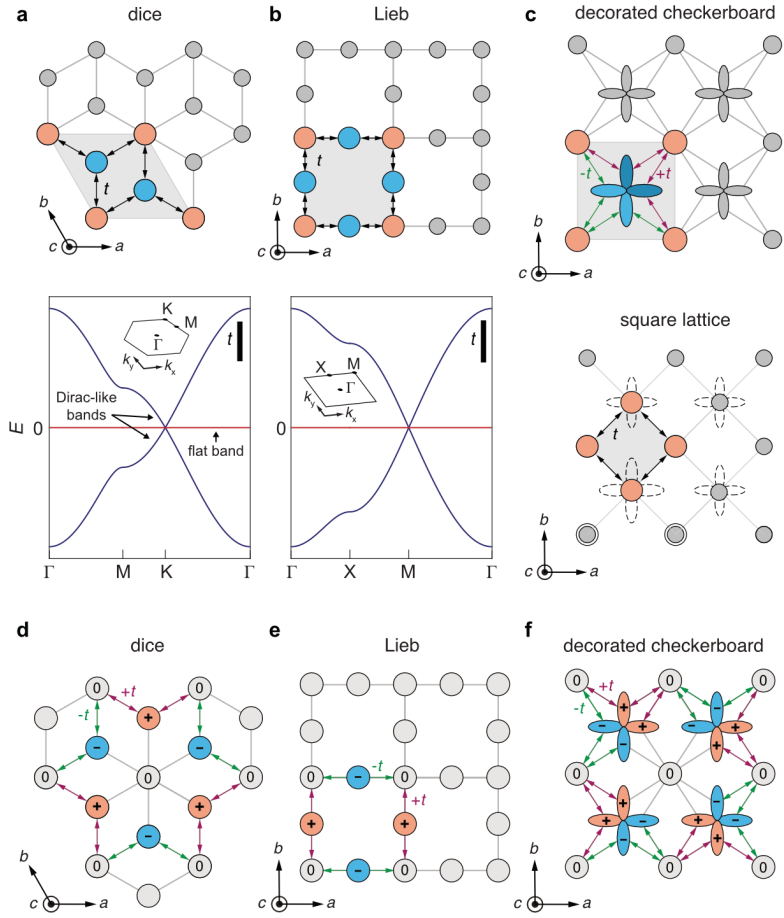

**Extended Data Figure 1: Dice and Lieb lattice models**

**a** (top) The hexagonal dice lattice is an imbalanced bipartite structure: it can be divided into two sublattices with an unequal number of sites, orange and blue, with one and two sites per unit cell, respectively. (bottom) The imbalanced bipartite structure leads to an electron-hole symmetric band structure with dispersive, Dirac-like bands intersected at $E = 0$ by a flat band. **b** (top) The square Lieb lattice is also an imbalanced bipartite lattice, resulting in (bottom) a near identical electronic structure to the dice model. **c** (top) Removing the two orthogonal orbitals (blue) from the central site of the decorated checkerboard reduces this system to the (bottom) primitive square Bravais lattice. **d** The compact localized state (CLS) for the dice lattice. Destructive interference of electrons hopping away from the hexagon shaped plaquette causes electrons to be localized. **e** An analogous CLS is found in the Lieb lattice, this time on a square plaquette. **f** The CLS in the decorated checkerboard model also has a square real-space structure. However, unlike the Lieb lattice, it has an additional $[d_{xz}, d_{yz}]$ orbital texture that ensures destructive interference of electrons hopping outside of the plaquette.

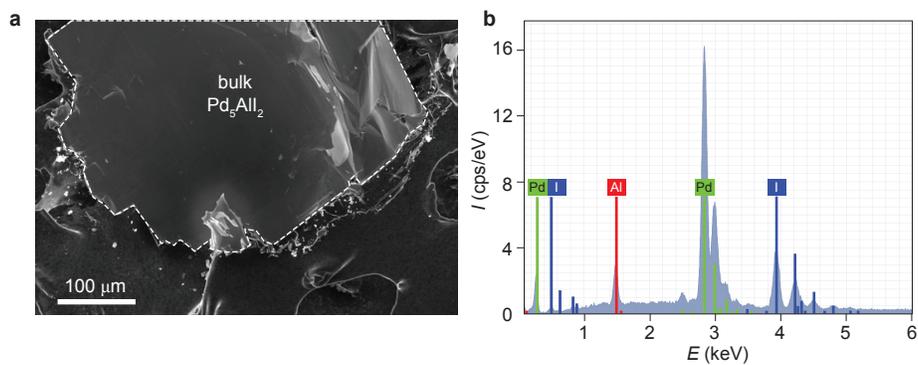

**Extended Data Figure 2: SEM imaging and EDS of Pd$_5$AlI$_2$**

**a** SEM image of Pd$_5$AlI$_2$ single crystal outlined in white. **b** EDS from a representative area scan on the crystal shown. Using statistics collected from ten area scans using the Al content as the reference, the Pd stoichiometry is $4.783 \pm 0.174$, Al stoichiometry is 1.0, and I stoichiometry is $1.843 \pm 0.049$.

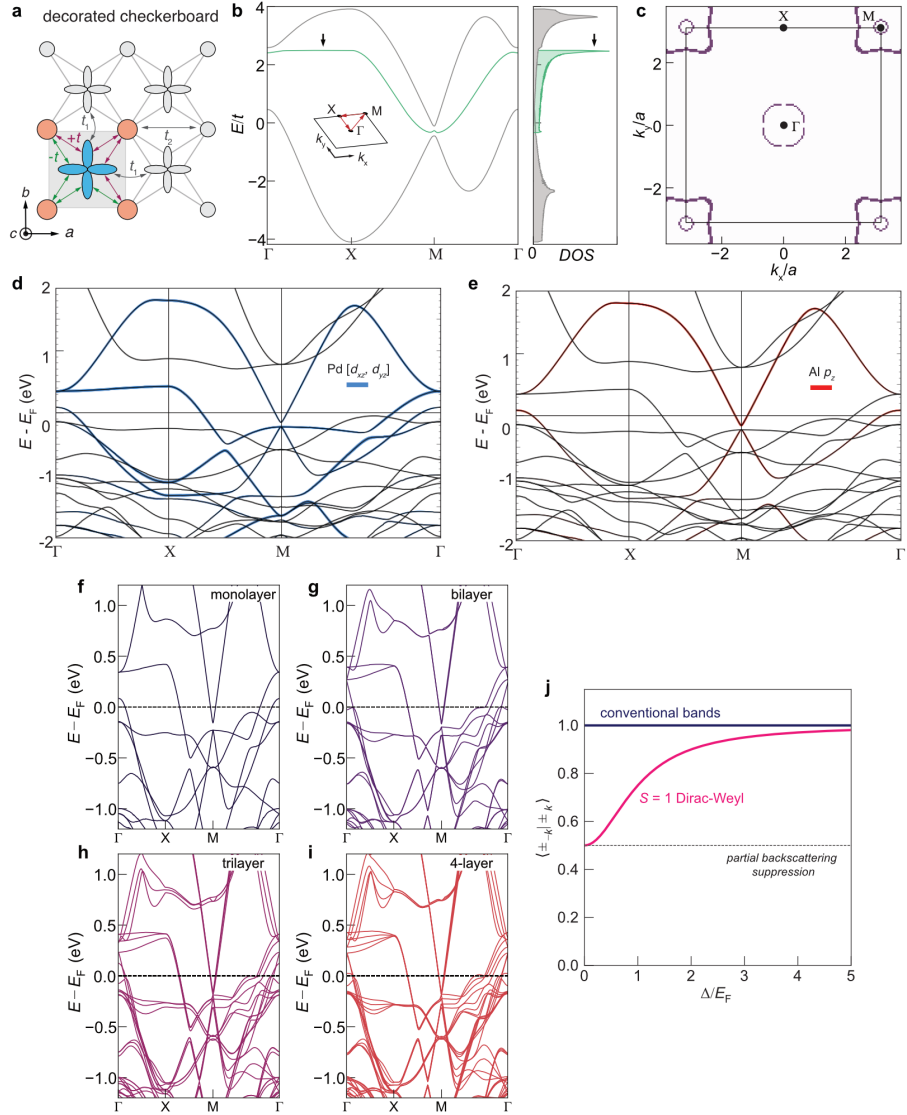

**Extended Data Figure 3: Decorated checkerboard model and suppressed backscattering**

**a** Decorated checkerboard model with direction- and orbital-dependent sign changes of the hopping energy $t$. Next-nearest neighbor hoppings $t_1$ and $t_2$ are also shown. **b** Band structure (left) and density of states (right) of the decorated checkerboard model including next-nearest neighbor, distinct on-site energies, and spin-orbit coupling. **c** Fermi surface from the extended model at $E = 0$, showing three distinct pockets. Projection of the DFT band structure onto **d** Pd $[d_{xz}, d_{yz}]$ and **e** Al $p_z$ orbitals of the PdAl checkerboard embedded within $Pd_5AlI_2$ monolayers. Bands near $E_F$ have significant contributions from these orbitals. DFT slab band structure calculations for **f** monolayer **g** bilayer **h** trilayer **i** and 4-layer thick $Pd_5AlI_2$ flakes. **j** Overlap between momentum-reversed eigenstates for the case of $S = 1$ DFs and conventional electrons. Backscattering can be partially suppressed for $S = 1$ DFs.

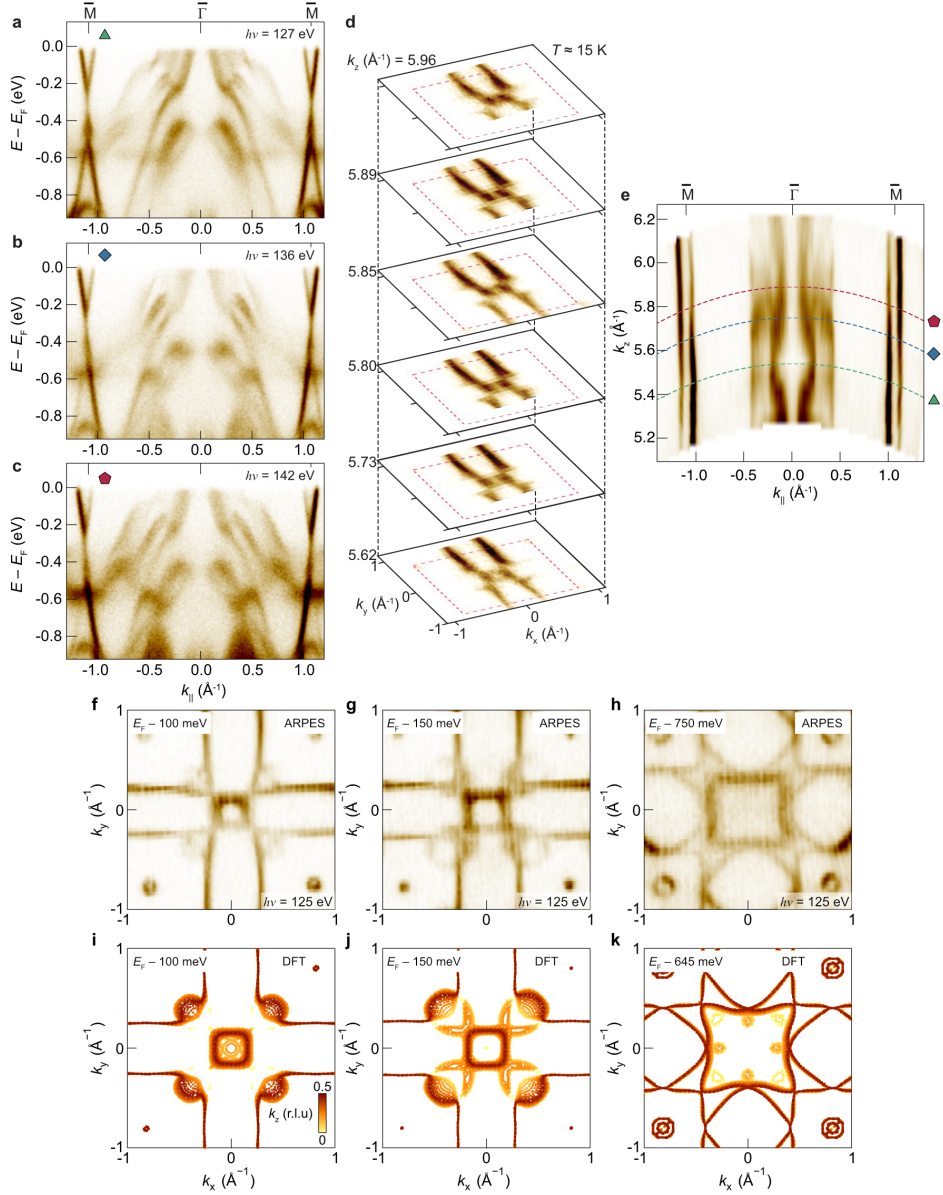

**Extended Data Figure 4: ARPES electronic structure and photon energy dependence**

ARPES $E(k)$ intensity map (LH polarization) along the $\overline{M} - \overline{\Gamma} - \overline{M}$ line of the surface BZ at various incoming photon energies $h\nu$, **a** 127 eV **b** 136 eV and **c** 142 eV, probing distinct slices of $k_z$, marked in **e**. The weak variation of these maps evidences the quasi-2D nature of the electronic structure. **d** ARPES Fermi surface cross-sections in the $k_x - k_y$ plane (LV polarization) at various $k_z$. The cross-sections are similarly invariant with $k_z$. **e** An ARPES $k_z - k_\parallel$ map along the $\overline{M} - \overline{\Gamma} - \overline{M}$ line (LH polarization) exhibits cylindrical FSs for all three pockets. The dashed lines show the approximate $k_z$ sampled by the $E(k)$ maps **a** – **c**. ARPES FS maps ($h\nu$ = 125 eV) at **f** 100 meV **g** 150 meV **h** and 750 meV below $E_F$. The experimental maps show good correspondence to DFT calculated FSs at **i** 100 meV **j** 150 meV and **k** 645 meV below $E_F$.

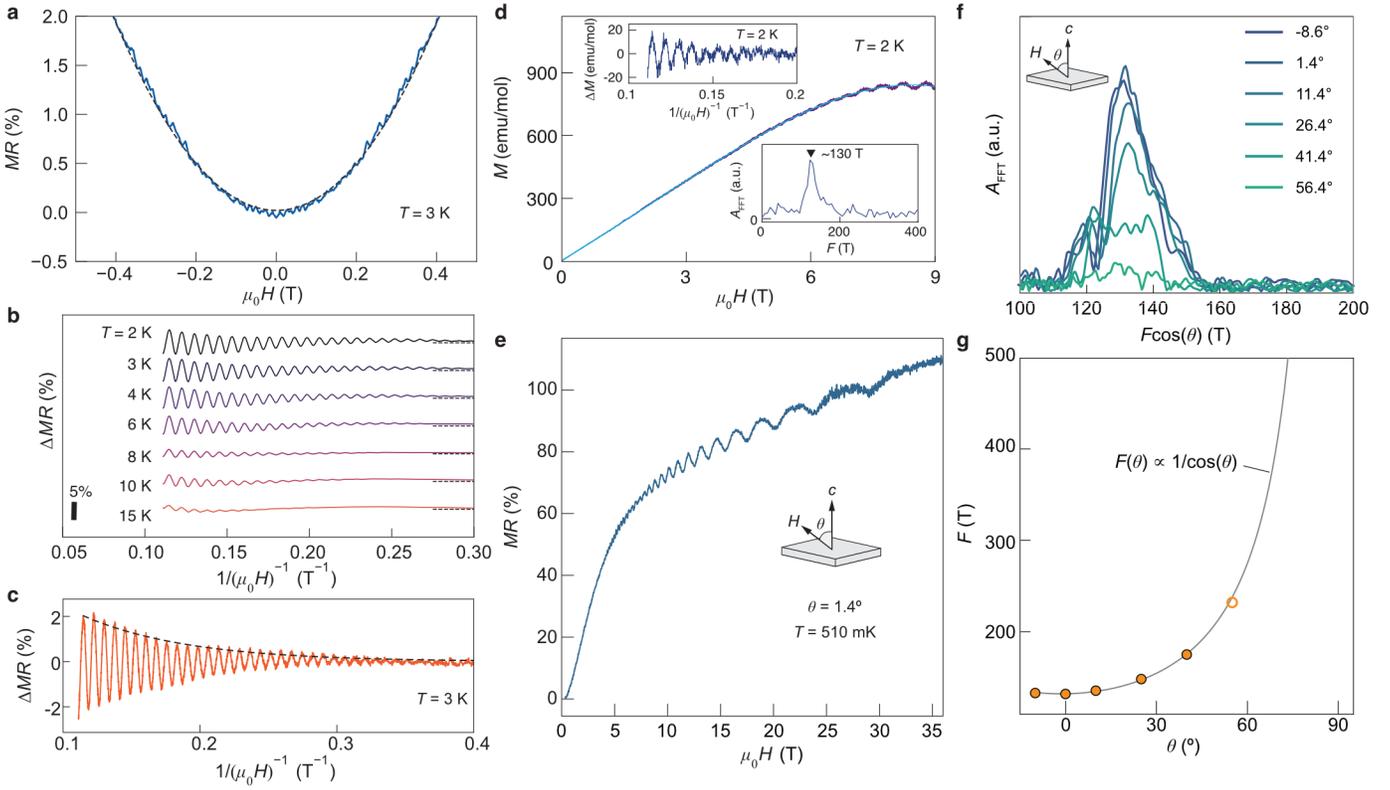

**Extended Data Figure 5: Bulk magnetoresistance and quantum oscillations**

**a** $MR(H)$ at low $H$ with a quadratic fit obtained by least-squares regression. **b** SdH oscillations $\Delta MR(1/H)$ at various fixed $T$ extracted by subtracting a 3$^{rd}$ order polynomial background from $MR(H)$. **c** We extract the quantum mobility of the $\alpha_M$ pocket carriers by fitting the exponential suppression of $\Delta MR(1/H)$. **d** Field-dependent magnetization $M(H)$. By removing a polynomial background (blue) we can isolate (inset, top) dHvA oscillations $\Delta M(1/H)$. (inset, bottom) FFT analysis of $\Delta M(1/H)$ reveals a peak at $F \approx 130$ T, in nominal agreement with SdH oscillations in bulk crystals and exfoliated flakes. **e** $MR(H)$ measured up to 36 T shows saturating behavior, and the absence of quantum oscillations from the $\beta_\Gamma$ and $\gamma_M$ pockets in this $H$ range. **f** Fast Fourier transport of SdH quantum oscillation at various fixed $\theta$ relative to the $c$-axis, plotted versus the scaled frequency $F\cos(\theta)$. **g** The peak frequency $F$ versus $\theta$ can be fit by the $1/\cos(\theta)$ dependence anticipated for quasi-2D Fermi surfaces. We show the frequency at $\theta = 56.4°$ as an open circle, given its comparatively weaker signature in the FFT.

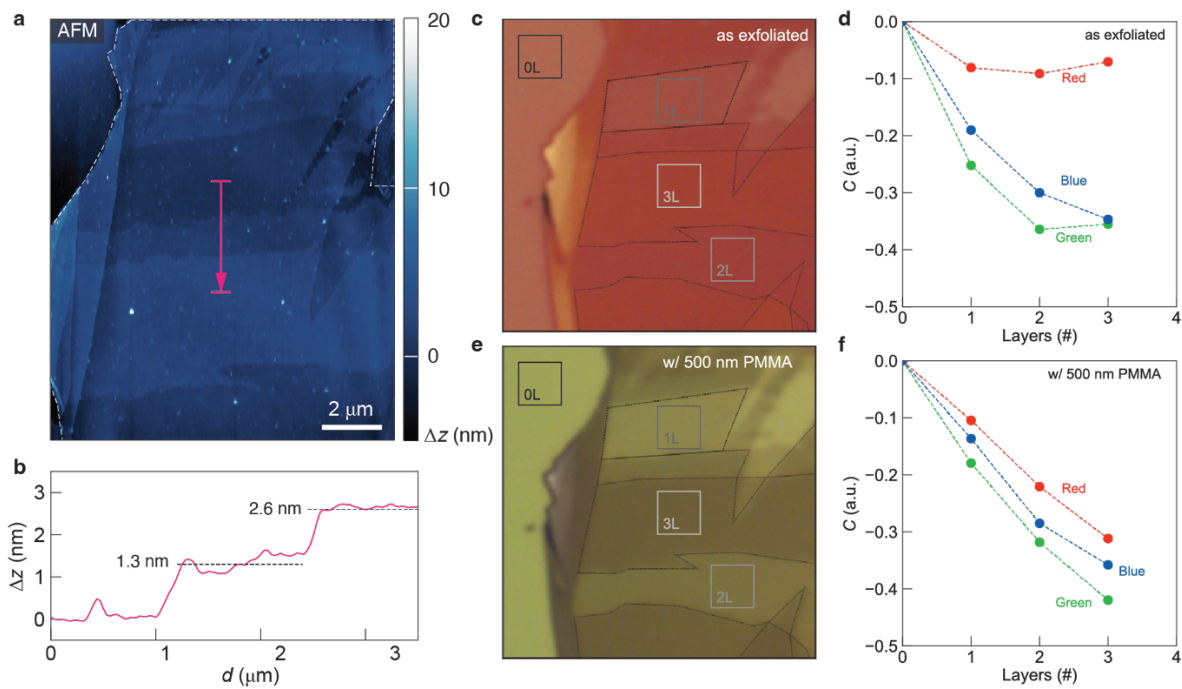

**Extended Data Figure 6: Pd$_5$AlI$_2$ AFM topography and optical contrast**

**a** Topography of terraced Pd$_5$AlI$_2$ flake exfoliated onto SiO$_2$/Si wafer measured by AFM with **b** linecut along magenta segment showing single Pd$_5$AlI$_2$ layer steps. **c** Optical microscope image of the same flake showing **d** contrast in the red, green, and blue color channels between regions of varying thickness. **e** Spinning a layer of PMMA over the flake **f** enhances optical contrast, making it easier to distinguish regions with different layer number.

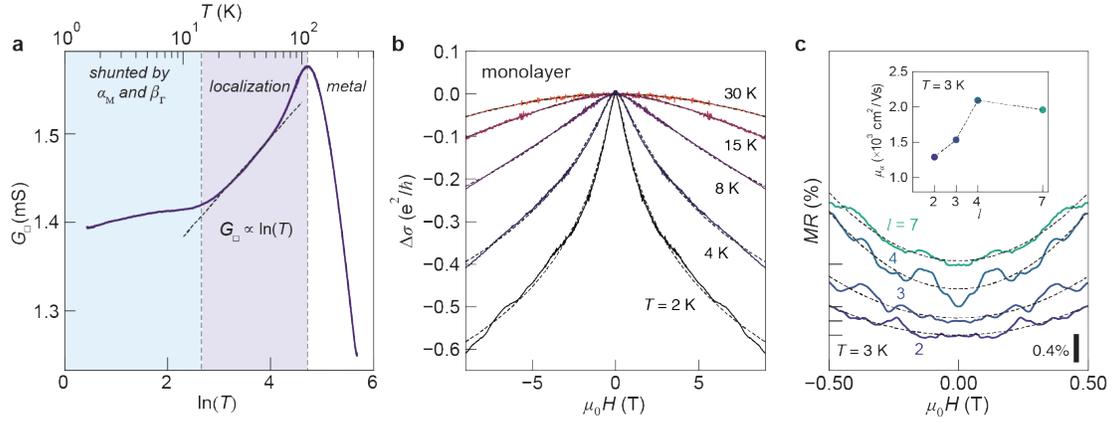

**Extended Data Figure 7: Device *T* and *H* dependent transport**

**a** Sheet conductance $G_\square$ versus $T$ for monolayer device showing localization of $\gamma_M$ below 100 K. For $T <$ 13 K we see a deviation from the localization behavior which we attribute to shunting by the still mobile $\alpha_M$ and $\beta_\Gamma$ carriers. **b** Magnetoconductance $\Delta\sigma(H)$ in the monolayer limit at various fixed $T$ follows the Hikami-Larkin-Nagaosa form (dashed lines) characteristic of weak anti-localization. **c** $MR(H)$ at low $H$ and $T = 3$ K for devices with various $l$. Quadratic fits obtained by least-squares regression are used to estimate (inset) the mobility of the $\alpha_M$ pocket.

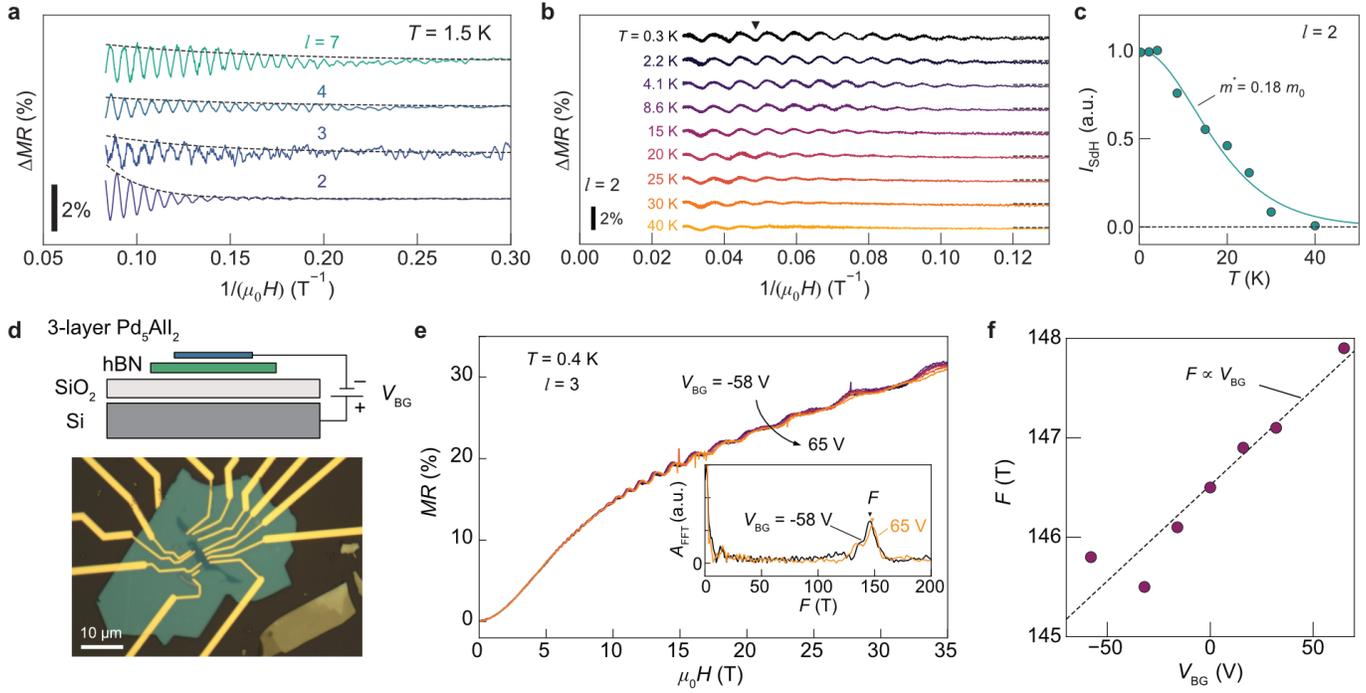

**Extended Data Figure 8: Device quantum oscillations analysis**

**a** We use the exponential suppression of $\Delta MR(1/H)$ at large $1/H$ to extract the quantum mobility $\mu_q$ of the $\alpha_M$ pocket for devices with various $l$. **b** $\Delta MR(1/H)$ at various fixed $T$ for a bilayer device ($l = 2$). **c** Lifshitz-Kosevich analysis of quantum oscillation amplitude extracted at $1/(\mu_0 H) = 0.05$ T$^{-1}$ (black triangle in **b**) these oscillations yields $m^* \approx 0.18\,m_0$, in good agreement with bulk SdH oscillations. **d** Schematic and optical microscope image of trilayer ($l = 3$) Pd$_5$AlI$_2$ transport device on hBN substrate. **e** Back gate voltage $V_{BG}$ dependence of high-field $MR(H)$ at $T = 0.4$ K. (inset) FFT spectra of the SdH oscillations shows that the oscillation frequency becomes larger with increasing $V_{BG}$. **f** SdH oscillation frequency $F(V_{BG})$ exhibiting electron-like behavior, consistent with the $\alpha_M$ pocket.

**Extended Data Table 1**: **Single crystal X-ray refinement**

Selected structure refinement data for $Pd_5AlI_2$.

| | |
|---|---|
| Empirical formula | $Pd_5AlI_2$ |
| Formula weight | 812.78 |
| Temperature | 101 K |
| Wavelength | 0.71073 Å |
| Crystal system | Tetragonal |
| Space group | I4/*mmm* |
| Unit cell dimensions | $a = 4.0342(1)$ Å, $\alpha = 90°$ |
| | $b = 4.0342(1)$ Å, $\beta = 90°$ |
| | $c = 19.4690(7)$ Å, $\gamma = 90°$ |
| Volume | 316.854(19) Å$^3$ |
| Z | 2 |
| Density (calculated) | 8.519 g/cm$^3$ |
| Absorption coefficient | 23.663 mm$^{-1}$ |
| $F(000)$ | 698 |
| Crystal size | $0.16 \times 0.14 \times 0.059$ mm$^3$ |
| $\theta$ range for data collection | 4.187 to 29.982° |
| Index ranges | $|h| \leq 5$ |
| | $|k| \leq 5$ |
| | $|l| \leq 26$ |
| Reflections collected | 1452 |
| Independent reflections | 175 [$R_{int} = 0.0394$] |
| Completeness to $\theta = 25.242°$ | 99.1% |
| Refinement method | Full-matrix least-squares on $F^2$ |
| Data / restraints / parameters | 175 / 0 / 13 |
| Goodness-of-fit | 1.224 |
| Final $R$ ins [$I > 2\sigma(I)$] | $R_{obs} = 0.0184$, $wR_{obs} = 0.0412$ |
| $R$ indices [all data] | $R_{all} = 0.0186$, $wR_{all} = 0.0413$ |
| Extinction coefficient | 0.0196(8) |
| Largest diff. peak and hole | 1.462 and $-1.994$ e·Å$^{-3}$ |

$R = \sum ||F_o| - |F_c|| / \sum |F_o|$, $wR = \left[ \sum w(|F_o|^2 - |F_c|^2)^2 / \sum w|F_o|^4 \right]^{1/2}$, and $w = 1 / \left[ \sigma^2 F_o^2 + (0.0198 P)^2 \right]$ with $P = (F_o^2 + 2F_c^2)/3$. $F_o$ and $F_c$ are the observed and calculated structure factors, respectively.

**Extended Data Table 2**: **Survey of exfoliated van der Waals metals**

Survey of 2D carrier density per layer $|n_{2D}|$ and $\mu$ across various exfoliated few-layer vdW metals and semimetals. Thickness of each sample in number of layers shown in parentheses.

| Sample | $|n_{2D}|$ (cm$^{-2}$) | $\mu$ (cm$^2$/Vs) | air stable? | Ref. |
|---|---|---|---|---|
| CHet-Ga (3L) | $1.9 \times 10^{15}$ | 51 | Yes | (45) |
| $H$-NbSe$_2$ (2L) | $1.3 \times 10^{15}$ | 64 | No | (77,78) |
| Pd$_5$AlI$_2$ (7L) | $1.1 \times 10^{15}$ | 2,600 | Yes | (this work) |
| Bi$_2$Sr$_2$CaCu$_2$O$_{8+\delta}$ (6L) | $6.7 \times 10^{14}$ | 12 | No | (79) |
| GdTe$_3$ (17L) | $3.8 \times 10^{14}$ | 5,700 | No | (81) |
| 1$T$'-WS$_2$ (160L) | $1.8 \times 10^{14}$ | 418 | Yes | (82) |
| FeSe (3L) | $7.4 \times 10^{13}$ | 331 | No | (80) |
| IG-MoS$_2$ (1L) | $8.0 \times 10^{13}$ | 813 | Yes | (83) |
| HP-Bi (25L) | $7.5 \times 10^{13}$ | 3,030 | Yes | (84) |
| Few-layer graphene (4L) | $1.0 \times 10^{13}$ | 100,000 | Yes | (37) |